\documentclass[bimj,fleqn]{w-art}
\usepackage{times}
\usepackage{mathrsfs}
\usepackage{w-thm}
\usepackage[authoryear]{natbib}
\theoremstyle{plain}

\theoremstyle{definition}

\usepackage[]{graphicx}
\chardef\bslash=`\\ 

\usepackage{algorithmic}
\usepackage{algorithm}
\usepackage{xcolor, soul}
\usepackage{amsfonts}
\usepackage{centernot}
\newcommand{\indip}{\mathrel{\perp\mspace{-10mu}\perp}}
\newcommand{\nindip}{\centernot{\indip}}
\DeclareMathOperator*{\argmin}{argmin}
\newcommand{\boldmcal}[1]{\boldsymbol{\mathcal{#1}}}
\usepackage{array}
\usepackage{hyperref}

\begin{document}
\Volume{52}
\Issue{61}
\Year{2010}
\pagespan{1}{}
\keywords{Function-on-function regression; Gaussian functional graphical models; Graphical models with covariates;  Group-based EEG connectivity; High-dimensional data.\\
}  

\title[Running title]{A neighbour selection approach for identifying differential networks in conditional functional graphical models}
\author[Alessia Mapelli {\it{et al.}}]{Alessia Mapelli\footnote{Corresponding author: {\sf{e-mail: alessia.mapelli@polimi.it}}}\inst{,1,2}} 
\address[\inst{1}]{MOX, Department of Mathematics, Politecnico di Milano, Piazza Leonardo da Vinci 32, Milano 20133, Italy}
\address[\inst{2}]{Health Data Science Research Centre, Human Technopole, Viale Rita Levi-Montalcini, 1, Milano, 20157, Italy}
\author[]{Laura Carini\inst{3}}
\address[\inst{3}]{Life Science Computational Laboratory, IRCCS Azienda Ospedaliera Metropolitana, Genova, Italy}
\author[]{Francesca Ieva\inst{1,2}} 
\author[]{Sara Sommariva\inst{4}}
\address[\inst{4}]{ Dipartimento di Matematica, Università degli Studi di Genova, via Dodecaneso 35, Genova, 16146, Italy}
\Receiveddate{zzz} \Reviseddate{zzz} \Accepteddate{zzz} 

\begin{abstract}

Estimating how different brain regions communicate with each other using EEG data is valuable both for medical research and clinical diagnosis. This involves quantifying the statistical dependencies among the activities of different brain areas, captured by the time-varying electric field recorded by scalp sensors. These dependencies can vary within and across individuals also in relationship with external factors such as age, mental state, or disease severity.

Motivated by this problem, we propose a novel neighbor selection approach based on Gaussian functional graphical models and functional-on-functional regression to identify which brain regions interact and how interaction strength changes with individual features or covariates (e.g., age or clinical status). Our approach is fully automated and data-driven, and, in principle, can handle any number of continuous and categorical covariates simultaneously. Unlike existing approaches, it also produces results that are easy to interpret: one can directly assess whether the strength of each estimated interaction increases or decreases as the value of a given covariate varies.

We evaluate our method through extensive simulation experiments and an application to real EEG data. The results demonstrate clear advantages over existing approaches, including more accurate estimation of brain connections and reduced computational cost, especially in high-dimensional settings involving a large number of brain regions and large sample sizes.

\end{abstract}

\maketitle                   







\section{Introduction}\label{sec-intro}
Estimating brain functional connectivity from electroencephalographic (EEG) data consists of quantifying the conditional dependencies between the activity of different brain areas from the time-varying electric field recorded by sensors placed on the scalp~\citep{sakkalis2011}. A large number of variables, such as age, mental status, and severity of neurodegenerative disorders, may influence these dependencies both within and across individuals~\citep{mueller2013,nentwich2020}. 
To provide a few motivating examples, EEG functional connectivity has been shown to vary with the clinical phenotype of early-stage patients with Dementia with Lewy Bodies (DLB) or Alzheimer's disease \citep{carini2025,hasoon2024}.
Specifically, DLB patients are usually grouped based on the presence of four clinical signs: visual hallucinations, Rapid Eye Movement (REM) sleep behavior disorder, cognitive fluctuation, and Parkinsonism. A precise characterization of differences in EEG functional connectivity between patients who present any of these signs and those who do not is still under investigation \citep{carini2025}. Similarly, a few recent studies have demonstrated that differences in EEG functional connectivity can discriminate alcohol and drug abusers from healthy controls \citep{khajehpour2019,mumtaz2018}. Hence, quantifying differences in functional connectivity across conditions (e.g., DLB patients with and without a given clinical sign, or alcohol abusers versus controls) may enable the identification of timely markers of these disorders.
Despite these results, the use of EEG functional connectivity in clinical settings remains limited, and the development of rigorous statistical techniques for robustly estimating the conditional dependency structure within the brain and its variation across conditions remains an open issue \citep{cao2022,sommariva2025,vallarino2020}. In this paper, we introduce a novel functional graphical modeling framework to overcome current limitations in the statistical analysis of EEG connectivity. By enabling the simultaneous estimation and comparison of connectivity networks across multiple conditions, the proposed methodology provides a principled approach for studying how brain dependency structures vary with clinical and demographic factors.

Graphical models \citep{lauritzen1996} have become a popular strategy in the last two decades for inferring conditional dependencies within the elements of a (high-dimensional) random vector. They consist of undirected graphs in which the nodes correspond to the elements of the random vector, and the edges encode pairwise conditional dependencies. When the random vector follows a multivariate Gaussian distribution, such edges correspond to the non-zero elements of its precision matrix. Functional graphical models \citep{qiao2019} extend the notion of graphical models to graphs whose nodes represent random functions, thus allowing the characterization of conditional dependencies within multivariate random functions. In particular, Gaussian functional graphical models deal with multivariate Gaussian processes. 

A typical workflow for estimating (Gaussian) functional graphical models consists of: (i) deriving a finite-dimensional representation of the observed functions by computing their projection scores with respect to a defined orthonormal basis; 
(ii) estimating a precision matrix from the computed projection scores to infer the conditional dependence graph. Two main families of approaches have been proposed for the second step~\citep{chen2024}. The first seeks to estimate the entire precision matrix directly by maximizing the corresponding likelihood under sparsity-inducing penalties, typically via group-Lasso regularization~\citep{qiao2019,zapata2022}. The second relies on neighbor selection schemes, where sparse functional regression models are employed to estimate the local neighborhood of each node separately~\citep{zhao2024}. This second family of approaches offers several practical and theoretical advantages in the settings considered here. By estimating the structure of the covariance operator one row at a time, it avoids the need to define the whole precision matrix, which may not be well-defined if the random functions lie in an infinite-dimensional space or if the corresponding covariance operator is not partially separable \citep{zapata2022}. Furthermore, since each node can be treated independently, these approaches efficiently scale to high-dimensional settings involving a large number of nodes and can exploit prior information, for example, by limiting the analysis to a subset of nodes of interest, as commonly done in seed-based functional connectivity studies.

Most existing methods in both families assume that all observed functions follow the same graphical models, without accounting for the effects of external covariates. Only a few recent works have focused on characterizing differences in the conditional dependence structures across populations. When comparing the graphical models associated with two populations or groups, the differential graph is commonly characterized by studying the support of the difference between their corresponding precision matrices \citep{zhao2014,xu2016,liu2014}. \cite{zhao2022} extended this idea to the functional setting and proposed an algorithm called FuDGE (Functional Differential Graph Estimation) to directly estimate the functional differential graph without explicitly computing the functional graphical model of the two populations. However, this method only accounts for the presence of one binary external variable. \cite{moysidis2021} proposed a method handling discrete covariates by assuming that all the subgroups in the population share a common graph structure with group-specific levels of sparsity. Hierarchical penalization is then used to infer the subgroups' precision matrices associated with the coefficients from the Karhunen-Loève expansion of the input functions. For handling both continuous and discrete external covariates, \cite{lee2023}
introduce a novel linear operator, the functional precision operator, that extends the notion of precision matrix and its estimation to the functional setting. However, also this approach is based on operator analysis, which can be computationally demanding and statistically inefficient in high-dimensional functional settings.

\begin{figure}
\centering{
\includegraphics[width=\textwidth]{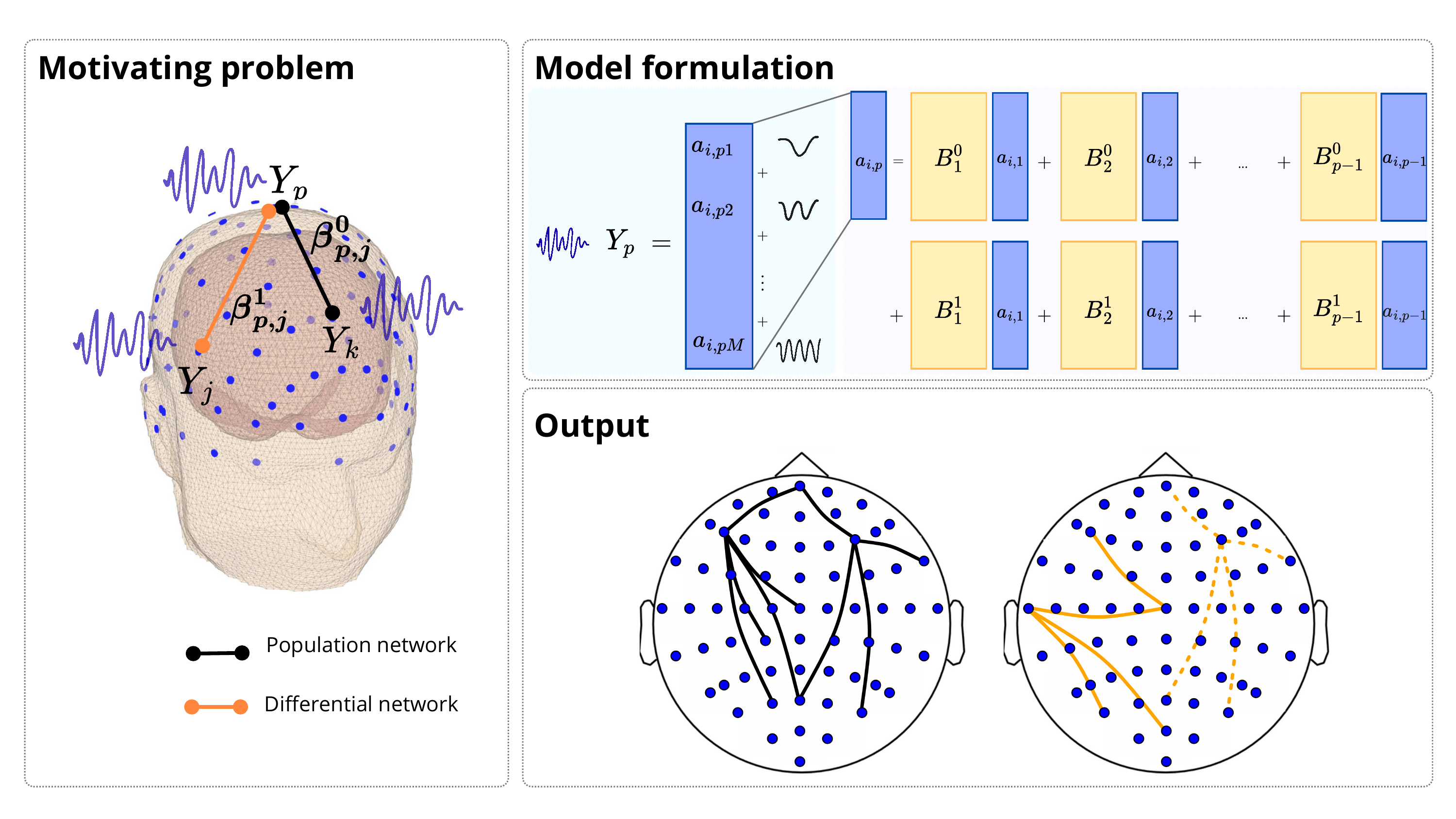}}\caption{Overview of the proposed approach. Motivating problem: we developed a neighbor selection approach based on function-on-function regressions to estimate the pairwise conditional dependencies across EEG recordings (population network) and how these dependencies vary with external covariates (differential networks). Model formulation: for each EEG sensor, by exploiting fPCA, the associated function-on-function regression model is transformed into a tractable vector-on-vector regression model. Output: exploiting the estimated regression coefficients, we assemble a population network (black), collecting pairwise dependencies shared by all samples, and a differential network for each external covariate (orange), collecting pairwise dependencies whose strength increases (plain lines) or decreases (dotted lines) when the value of the external covariate varies.}\label{fig:approach_overview}
\end{figure}

Here, we propose a neighbor selection strategy, schematically illustrated in Fig.~\ref{fig:approach_overview}, that extends the approach proposed by \cite{zhao2024} to the estimation of conditional functional graphical models whose structure may vary with an arbitrary number of continuous and discrete external covariates. In detail, for each node, we formulate the identification of its neighbors as a function-on-function (fof) regression problem, modeling the interaction between the response variables (in our case, the EEG time-series) and the covariates. The fof regression model is then approximated by projecting the observed functions in a node-specific basis determined via Functional Principal Component Analysis (FPCA). The coefficient matrices of the resulting vector-on-vector (vov) regression model are estimated through a group-lasso-penalized least squares approach. The result is a fully automated, data-driven approach that, given a set of observed functions, characterizes their conditional dependence structure by directly identifying the pairwise interactions that vary with each covariate. We demonstrate the benefit of our proposed approach in the framework of functional EEG connectivity by comparing its performance with that of FuDGE and three related variants on both simulated data and experimental EEG recordings from a public dataset concerning alcohol use disorder \citep{ingber1997,zhang1995}. The proposed framework is general and applicable across a broad range of settings and applications. An open-source implementation is publicly available at \texttt{https://github.com/AlessiaMapelli/condFGM}.

The paper is organized as follows. In Section \ref{sec:stat_model}, we described the proposed fof-regression-based neighbor selection approach. In Section \ref{sec:estimation}, we provide details on the estimation procedure for the coefficient matrices of the resulting approximated vov regression models. Section \ref{sec:results_simulation}
and Section \ref{sec:results_experdata} collects the results from the analysis of the simulated and the experimental EEG data, respectively. Our conclusions are offered in Section \ref{sec:discussion}.

\section{Stastistical model}\label{sec:stat_model}
In this section, we first introduce the notion of conditional functional graphical models, then we define the fof regression model associated with each node of the graphical model. An overview of the main definitions introduced in this section and the corresponding interpretation within our motivating example concerning the study of EEG functional connectivity is provided in Table \ref{tab:defs}.

\subsection{Conditional functional graphical model}

Let $(\Omega, \mathcal{F}, \mathbb{P})$ be a probability space and let $\mathcal{H} \subseteq \mathcal{L}^2(\mathcal{T})$ be a Hilbert subspace of the space of square-integrable real-valued functions $\mathcal{L}^2(\mathcal{T})$, where $\mathcal{T} \subseteq \mathbb{R}$ is a closed interval. 

Given $p>0$, we consider a Multivariate Gaussian Process (MGP) $\mathbf{Y}: \Omega \mapsto \mathcal{H}^p$ that we assume to be zero-mean. For all $\omega \in \Omega$ and $t \in \mathcal{T}$ we denote $\mathbf{Y}(\omega)(t) = (Y_1(\omega, t), \ldots, Y_p(\omega, t))$, were $Y_j(\omega, \cdot) \in \mathcal{H}$ for all $j = 1, \dots, p$. Throughout the paper, we will omit the dependence on $\omega$ and/or $t$ whenever this does not lead to ambiguity. In our application, $p$ denotes the number of EEG sensors, and $Y_j$ denotes the signal at the $j$-th sensor.

For each pair $(Y_j, Y_l)$, $j,l = 1, \dots, p$, the conditional cross-covariance function is defined as \citep{qiao2019}
\begin{displaymath}
C_{jl}(t',t)=Cov(Y_j(t'), Y_l(t)|\mathbf{Y}_{-\{j,l\}})
\end{displaymath}
where $\mathbf{Y}_{-\{j,l\}} = \left\{Y_k : k \in \{1, \dots, p \} \setminus \{j,l  \}\right\}$ and $(t', t) \in \mathcal{T}^2$.  We say that $Y_j$ and $Y_l$ are conditionally independent, and we write $( Y_j(t) \indip Y_l(t)  \mid \mathbf{Y}_{-\{j,l\}})$, if and only if $C_{jl}(t',t)=0$ for all $(t', t) \in \mathcal{T}^2$. Defined the set of vertices $V=\{1, \dots, p\}$, we aimed at characterizing the undirected graph $G=(V, E)$ where the set of edges 
\begin{equation}\label{eq:edges_FGM}
E = \{(j,l) \in V^2:  Y_j(t)  \nindip Y_l(t)  \mid \mathbf{Y}_{-\{j,l\}} \} 
\end{equation}
collects all the pairwise conditional interdependence within $\mathbf{Y}$.

In many practical scenarios, the graph $G$ may vary with external variables, such as patients' age or disease stage. Inspired by \cite{lee2023}, we used a conditional functional graphical model to account for this influence. In detail, given a $q$-dimensional random vector of covariates $\mathbf{X} = \left(X_1, \dots, X_q \right)$, for all $\mathbf{x} \in \mathbb{R}^q$ we assume that $\mathbf{Y} \mid \mathbf{X}=\mathbf{x}$ is a MGP following a graphical model defined by the graph $G(\mathbf{x})=\left(V, E(\mathbf{x})\right)$ being 
\begin{equation}\label{eq:edges_cond_graph}
E(\mathbf{x}) = \left\{(j,l) \in V^2:  Y_j(t)  \nindip Y_l(t)  \mid \left[\mathbf{Y}_{-\{j,l\}}, \mathbf{X}=\mathbf{x} \right]\right\} \, .
\end{equation}

\subsection{Function-on-function regression}
Given the functional graphical model defined by the graph $G$ in Eq. (\ref{eq:edges_FGM}), under mild assumptions \citep{zhao2024}, for each node $j \in V$ there exist $\{\beta_{j,k}(t, t')\}_{j\neq k}$ such that $||\beta_{j,k}||_{HS} < \infty$ and
\begin{equation}\label{eq:fof_regr_model}
Y_j(t) = \sum_{k \neq j} \int_\mathcal{T} \beta_{j,k}(t,t')Y_k(t')dt' + e_j (t) \, 
\end{equation}
being $e_j (\cdot) \indip Y_k(\cdot)$ for all $k\neq j$, and the set of neighbours of node $j$ in graph $G$ is given by  
\begin{displaymath}
\mathcal{N}_j :=\left\{k \in V  \setminus \{j\}: (j,k) \in E \right\} = \left\{k \in V \setminus \{j\} : ||\beta_{j,k}||_{HS} > 0 \right\} \, . 
\end{displaymath}
To account for the influence of the external covariates, we assume that the regression coefficients in (\ref{eq:fof_regr_model}) depend linearly on the component of $\mathbf{X}$. Hence for all $c=0, \dots, q$, there exists $\{ \beta_{j,k}^c(t, t') \}_{j\neq k}$ such that
\begin{equation}\label{eq:cond_coeffs}
\beta_{j,k}(t, t') = \beta^0_{j,k}(t, t') + \sum_{c=1}^q \beta^c_{j,k}(t, t') \, X_c = \sum_{c=0}^q \beta^c_{j,k}(t, t') \, X_c
\end{equation}
where we set $X_0 = 1$. By incorporating (\ref{eq:cond_coeffs}) into the model in (\ref{eq:fof_regr_model}) we obtain
\begin{eqnarray}
Y_j(t) & = & \sum_{k \neq j} \int_\mathcal{T} \beta^0_{j,k}(t,t')Y_k(t')dt' + \sum_{c=1}^q \sum_{k \neq j} \int_\mathcal{T} \beta^c_{j,k}(t,t') X_c Y_k(t')dt'+ e_j (t) \nonumber \\
& = & \sum_{c=0}^q \sum_{k \neq j} \int_\mathcal{T} \beta^c_{j,k}(t,t') X_c Y_k(t')dt'+ e_j (t) \label{eq:fof_cond_regr_model} 
\end{eqnarray}
The complete set of neighbours of node $j$ according to the conditional graph $G(\mathbf{x})$ in Eq. (\ref{eq:edges_cond_graph}) will be given by 
\begin{displaymath}
\mathcal{N}_j (\mathbf{x}) :=\left\{k \in V  \setminus \{j\}: (j,k) \in E(\mathbf{x}) \right\} = \left\{k \in V \setminus \{j\} : ||\sum_{c=0}^q\beta^c_{j,k}X_c||_{HS} > 0 \right\} \, . 
\end{displaymath}

Our method aims at characterizing how the conditional dependence structure of $\mathbf{Y}$ varies with each covariate, assuming the set of vertices $V = \{1, \dots, p\}$ to be fixed across samples. In model~(\ref{eq:fof_cond_regr_model}), the baseline coefficients $\beta^0_{jk}(t,t')$ encode the population-level pairwise conditional interdependencies shared by all samples, while $\beta^c_{jk}(t,t')$ captures how such
relationships vary with the $c$-th covariate. This enables the identification, for each covariate $ X_c$, of the edges $(j,k)$ that differ for different values of the covariate from the population-level. To this end, for each $c=0,\ldots,q$, we define the covariate-specific undirected graph $G^c=(V,E^c)$ such that the associated set of neighbours of node $j$ is given by 
\begin{equation}\label{eq:def_diff_nodesset}
\mathcal{N}_j^c=\left\{k\in V\setminus\{j\}: (j,k) \in E^c \right\} = \left\{k\in V\setminus\{j\}: ||\beta^c_{j,k}\|_{HS}>0\right\} \, .
\end{equation}
Here, $G^0$ represents the population-level (baseline) network, whereas for $c\ge 1$ the graph $G^c$ encodes a differential network collecting the pairs of nodes for which the conditional association varies with the $c$-th covariate $X_c$. For any covariate value $\mathbf{x}\in\mathbb{R}^q$, the conditional model $G(\mathbf{x})$ in Eq. (\ref{eq:edges_cond_graph}) is obtained by
combining the baseline and covariate components through the linear decomposition
$\beta_{jk}(t,t';\mathbf{x})=\beta^0_{jk}(t,t')+\sum_{c=1}^q x_c\,\beta^c_{jk}(t,t')$. Thus, an edge may be present in $G(\mathbf{x})$ due to the baseline term, due to a covariate-linked term, or due to their combination, while the graphs $\{G^c\}_{c=1}^q$ isolate which
edges are differential with respect to each specific covariate. This explicit decomposition into a population network and covariate-specific differential networks, and their simultaneous estimation within a unified model, is the key novelty of our approach. \\
\noindent \begin{remark} The fof model in Eq.  (\ref{eq:fof_cond_regr_model}) easily accounts for both continuos and categorical covariates. In particular, a categorical covariate with $\ell$ categories can be incorporated in the model by fixing a reference category and by defining $\ell-1$ binary variables such that
\begin{equation}
X_c = \left\{\begin{array}{cc}  
1 & \text{if an observation belongs to category $c$}\\
0 & \text{otherwise}
\end{array} \right. \quad c = 1, \dots, \ell-1 \, .
\label{eq:one_hot_enc}
\end{equation}
\end{remark}
For an observation in the reference group (i.e., $X_c = 0$ for all $c = 1, \dots, \ell-1$) it then holds
\begin{displaymath}
Y_j(t) = \sum_{k \neq j} \int_\mathcal{T} \beta^0_{j,k}(t,t') Y_k(t')dt'+ e_j (t)  \, ,
\end{displaymath}
while for an observation in the $c$-th group (i.e., $X_c = 1$), $c=1, \dots, \ell-1$, we can write
\begin{displaymath}
Y_j(t) = \sum_{k \neq j} \int_\mathcal{T} (\beta^0_{j,k}(t,t') + \beta^c_{j,k}(t,t'))Y_k(t')dt'+ e_j (t)  \, .
\end{displaymath}
In this case, the population network $G^0 = (V, E^0)$ collects the conditional dependence relationships within the reference group, while the differential network $G^c = (V, E^c)$ represents the pairwise relationship in which the $c$-th group differs from the reference one.

\section{Estimation procedure}\label{sec:estimation}
Suppose we observe $n$ functions $\{ \mathbf{y}_i (\cdot) := \left( y_{i,1}(\cdot), \dots, y_{i,p}(\cdot) \right)\}_{i=1}^n$ and the corresponding covariates $\{ \mathbf{x}_i := \left( x_{i,1}, \dots, x_{i,q} \right)\}_{i=1}^n$ which are realization of $n$ i.i.d. random copies of $\mathbf{Y}$ and $\mathbf{X}$, respectively. In this paper, we present a neighbor selection procedure for estimating both the population network $G^0 = (V, E^0)$ and the differential networks $G^c = (V, E^c)$, $c=1, \dots, q$. Hence, for each node $j=1, \dots, p$ we seek to estimate the sets of nodes $\mathcal{N}_j^c$, $c=0, \dots, q$ defined by (\ref{eq:def_diff_nodesset}). In this Section, details of the estimation procedure for the fof regression model presented in (\ref{eq:fof_cond_regr_model}) will be provided, along with practical implementation details for the proposed approach. Since the neighbor selection is carried out independently for each target node, for the ease of notation in the following, we will focus on the last node, i.e., $j=p$.

\subsection{Vector-on-vector regression}
The fof regression model in (\ref{eq:fof_cond_regr_model}) can be approximated in a tractable finite-dimensional vector-on-vector (vov) linear regression problem as follows. Let $\{ \phi_{m}^{(p)}\}_{m=1}^\infty$ be an orthonormal basis of $\mathcal{H}$, and let $M > 0$ be a positive constant. The actual choice of the basis, which is defined independently for each node, is discussed in Appendix \ref{sec:basis_def}. Each observed function is approximated with the corresponding projection onto the first $M$ elements of the basis, namely 
\begin{eqnarray}
y_{i,k} &  = & \sum_{m=1}^\infty a_{i,km}^{(p)} \phi_{m}^{(p)} \label{eq:serie_Y} \\
 & \simeq & \sum_{m=1}^M a_{i,km}^{(p)} \phi_{m}^{(p)} \label{eq:approx_serie_Y}
\end{eqnarray}
where, for all $k=1, \dots, p$ and $i=1, \dots, n$, $a_{i,km}^{(p)} = <y_{i,k}, \phi_{m}^{(p)}>$ are the projection scores.

For the ease of notation, in the following, we will omit the dependence on the index $p$ of the projection scores and of the elements of the basis. 

Using Eq. (\ref{eq:approx_serie_Y}) and the fof model in (\ref{eq:fof_cond_regr_model}), it follows that for each observation $i=1, \dots, n$ the following representation holds:

\begin{equation}\label{eq:vov_regr_model}
\mathbf{a}_{i,p} = \sum_{c=0}^q\sum_{k=1}^{p-1}\mathbf{B}_{k}^c x_{i,c}\mathbf{a}_{i,k} + \boldsymbol{\rho}_{i} + \boldsymbol{\varepsilon}_{i} \, .
\end{equation}

In the previous model, $\mathbf{a}_{i,k} = ( a_{i,k1}, ...., a_{i,kM})^{\top}$ collects the first $M$ projection scores for the $k$-th node; $\mathbf{B}_{k}^c = (b^c_{k,mm'})_{1 \leq m,m' \leq M} \in \mathbb{R}^{M \times M}$ are regression coefficient matrices whose relationship with the parameters of the original fof regression model is given by

\begin{equation}
b^c_{k,mm'} = \int_{\mathcal{T}} \int_{\mathcal{T}} \beta^c_{p,k}(t,t')\phi_m(t)\phi_{m'}(t')dtdt' \quad \forall m,m' \geq 1 \, \;\label{eq:vov_fof_coef_corr}
\end{equation}
$\boldsymbol{\rho}_{i} = \left(\rho_{i,1}, \dots, \rho_{i,M} \right)$ is a bias term arising when the approximation in (\ref{eq:approx_serie_Y}) is employed with $\rho_{i,m} = \sum_{c=0}^q \sum_{k=1}^{p-1}\sum_{m'=M+1}^\infty b^c_{k,mm'} a_{i,km'}x_{i,c}$; and $\boldsymbol{\varepsilon}_{i} = \left(\varepsilon_{i,1}, \dots, \varepsilon_{i,M} \right)$, being $\varepsilon_{i,m} = \int_\mathcal{T} e_{i,p}(t) \phi_m(t) \, dt$, is a vector of model errors related to the error term $e_{i,p}(\cdot)$ in (\ref{eq:fof_cond_regr_model}).

According to the vov model in (\ref{eq:vov_regr_model})  and the coefficient correspondence in (~\ref{eq:vov_fof_coef_corr}), if $M$ is large enough, the set of neighbours of node $p$ in the differential graph $G^c$ associated to the $c$-th covariate can be approximated as
\begin{displaymath}
\mathcal{N}_p^c \simeq \left\{k \in V \setminus \{p\} : ||\mathbf{B}_k^c||_{F} > 0 \right\} \, .
\end{displaymath}

\begin{table}[htb]
\begin{center}
\caption{List of the main definitions introduced in the paper and corresponding interpretation in the context of EEG functional connectivity analysis.}\label{tab:defs}
\begin{tabular}{c>{\centering\arraybackslash}p{5cm}p{6cm}}
\hline
\textbf{Symbol} & \textbf{Mathematical Object} & \textbf{Interpretation}\\
\hline
$\mathbf{Y}(\cdot) = (Y_1(\cdot), \dots, Y_p(\cdot))$ & Multivariate Gaussian Process & EEG signal from $p$ sensors \\
$\mathbf{X} = (X_1, \dots, X_q)$ & Random vector & $q$ external covariates (e.g. group membership, age)\\
$\beta_{j,k}^c(\cdot, \cdot)$ & Fof regression coefficients &  Influence of sensor $k$ on sensor $j$ related to the $c$-th covariate \\
$G^0 = (V, E^0)$ & Population graphical model & Functional connectivity network shared by all samples\\
$G^c = (V, E^c)$ & Differential graphical model & Functional connectivity network varying with the $c$-th covariate\\
$\mathcal{N}^c_j$ & Set of neighbours & Signal influencing the $j$-th EEG signal according to $G^c$ $(c=0, \dots, q)$\\
$\mathbf{a}_{i,k}^{(j)} = (a_{i,k1}^{(j)}, \dots, a_{i,kM}^{(j)})$ & Projection scores & Signal from sensor $k$ projected on the node-specific basis $\{ \phi_m^{(j)} \}_{m=1}^M$ (dependency on $j$ often omitted) \\
$\mathbf{B}_{k}^c, \quad \widehat{\mathbf{B}}_{k}^c$ & Vov regression coefficient matrices and corresponding estimates & Influence of sensor $k$ on sensor $j$ related to the $c$-th covariate  \\
\hline
\end{tabular}
\end{center}
\end{table}

\subsection{Penalized least square estimation of the regression coefficient and of the differential networks}

Inspired by \cite{zhao2024} we compute a penalized least square estimator of the coefficient matrices within the vov regression model in Eq. (\ref{eq:vov_regr_model}). Specifically, given $n$ i.i.d samples $\{ (\mathbf{y}_i(\cdot), \mathbf{x}_i) \}_{i=1}^n$ we determine
\begin{equation}\label{eq:pen_least_square}
\begin{split}
 & \widehat{\mathbf{B}}_1^0, \dots, \widehat{\mathbf{B}}_{p-1}^0, \widehat{\mathbf{B}}_1^1, \dots, \widehat{\mathbf{B}}_{p-1}^1, \dots, \widehat{\mathbf{B}}_1^q, \dots, \widehat{\mathbf{B}}_{p-1}^q \in  \\
 & \quad \argmin_{\mathbf{B}_1^0, \dots, \mathbf{B}_{p-1}^0, \dots, \mathbf{B}_1^q, \dots, \mathbf{B}_{p-1}^q \in \mathbb{R}^{M \times M}} \left\{ \frac{1}{2n} \sum_{i=1}^n \left| \left| \mathbf{a}_{i,p} - \sum_{c=0}^q\sum_{k=1}^{p-1}\mathbf{B}_{k}^c x_{i,c}\mathbf{a}_{i,k} \right| \right|_2^2 + \lambda_p \sum_{c=0}^q\sum_{k=1}^{p-1} \left| \left| \mathbf{B}_{k}^c \right| \right|_F \right\}
    \end{split}
\end{equation}
where $\lambda_p$ is a regularization parameter whose value may depend on the node whose neighbors' set is being computed. In a reformulated form, specified in Appendix~\ref{supp:alg_details}, the optimization problem can be solved via the Alternating Direction Method of Multipliers (ADMM, \cite{Boyd2011}).

Once the estimates $\widehat{\mathbf{B}}_1^0, \dots, \widehat{\mathbf{B}}_{p-1}^0, \dots, \widehat{\mathbf{B}}_1^q, \dots, \widehat{\mathbf{B}}_{p-1}^q$ of the regression coefficients have been computed, the estimated set of neighbours of node $p$ within the differential graph $G^c$ is given by
\begin{equation} \label{eq:N_def_thr}
\widehat{\mathcal{N}}_p^c = \left\{k \in V \setminus \{p\} : ||\widehat{\mathbf{B}}_k^c||_{F} > \varepsilon_p \right\} \, ,
\end{equation}
where $\varepsilon_p > 0$ is a threshold parameter to be fixed.
 
The same procedure is carried out independently for each target node $j\in V$, yielding node-wise neighbourhood estimates $\widehat{\mathcal{N}}_j^c$, $c=0,\ldots,q$. Following the neighbourhood selection literature \citep{Meinshausen2006}, we recover an undirected edge set via OR- or AND-symmetrization:
\begin{equation}\label{eq:or_symm}
\widehat{E}^{c,\texttt{OR}} = \left\{(j, l) \in V^2 : l \in \widehat{\mathcal{N}}_j^c \ \ \texttt{OR} \ \  j \in \widehat{\mathcal{N}}_l^c  \right\},
\end{equation}
or, for a more conservative choice,
\begin{equation}\label{eq:and_symm}
\widehat{E}^{c,\texttt{AND}} = \left\{(j, l) \in V^2 : l \in \widehat{\mathcal{N}}_j^c \ \ \texttt{AND} \ \  j \in \widehat{\mathcal{N}}_l^c  \right\}.
\end{equation}

\paragraph{Characterizing the direction of change.}
The edge sets in~(\ref{eq:or_symm})--(\ref{eq:and_symm}) identify pairs of nodes whose conditional interdependencies varies with the $c$-th covariate. In contrast to approaches that only recover the support of a differential network, our formulation also allows to discern whether the interaction strength associated with each edge of this differential network increases or decreases as the covariate value varies. \\
Once $\widehat{\mathcal{N}}_p^c$ is obtained from (\ref{eq:N_def_thr}), the variation of the conditional interdependency between node $p$ and another node $k$ related the $c$-th covariate can be further specified by comparing the baseline block $\widehat{\mathbf{B}}_k^0$ with its covariate-specific counterpart $\widehat{\mathbf{B}}_k^c$. For each $k \in \widehat{\mathcal{N}}_p^c$, we define the Variation Rate (VR)
\begin{equation}\label{eq:relative_eff}
VR^c_{kp}= 
\frac{||\widehat{\mathbf{B}}_k^0 + \widehat{\mathbf{B}}_k^c||_{F}}{||\widehat{\mathbf{B}}_k^0||_{F}} \, .
\end{equation}
Values $VR^c_{kp}<1$ indicate an attenuation in the baseline association between nodes $p$ and $k$ for larger values of the covariate, whereas $VR^c_{kp}>1$ indicates a strengthening. For visualization and interpretability, one may alternatively report $\log(VR_{kp}^c)$, which is symmetric around zero (negative values correspond to reductions and positive values to increases).

These variations can be summarized for each differential edge $(k,p)$ in $G^c$ into a single weight as follows:
\begin{equation}\label{eq:edge_weights}
w_{kp}^c = \left\{ \begin{array}{lc}
VR^c_{kp} & \text{ if} \quad k \in \mathcal{N}_{p}^c \quad \text{and} \quad p \notin \mathcal{N}_k^c  \\
VR^c_{pk} & \text{ if} \quad k \notin \mathcal{N}_{p}^c \quad \text{and} \quad p \in \mathcal{N}_k^c \\
\sqrt{VR^c_{kp} \cdot VR^c_{pk} } & \text{ if} \quad k \in \mathcal{N}_{p}^c \quad \text{and} \quad p \in \mathcal{N}_k^c
\end{array} \right. \, .
\end{equation}



\subsection{Pipeline overview}
A schematic overview of the implemented pipeline is provided in Algorithm \ref{alg:pipeline}. In particular, we observe that in many practical scenarios, including the estimation of EEG functional connectivity, the functions $\{ \mathbf{y}_i \}_{i=1}^n$ are only observed on a discrete set of time-points $\{t_{\ell}\}_{\ell=1}^{L_i}$. Therefore, a smoothing step (Step~\ref{step:smoothing} of Algorithm\ref{alg:pipeline}) is required before applying the proposed approach. Details on how to perform this step are provided in Appendix \ref{sec:basis_def}, where we also discuss the choice of the basis for the finite-dimensional embedding (Step~\ref{step:dim_red}-a), providing practical guidance for selecting the truncation level $M$. Details on the hyperparameter selection procedure, within Step~\ref{step:dim_red}-b, can be found in Appendix \ref{sec:hyperparameters}: the group-lasso neighbourhood regressions are solved via an ADMM routine, and the node-specific tuning parameters $(\lambda_j^\ast,\epsilon_j^\ast)$ are selected automatically using selective cross-validation (SCV) with a BIC-type penalty, with an optional randomized search to reduce computational cost.\\

\begin{algorithm}
\caption{Neighbor selection approach for estimating population and differential functional graphical models.}\label{alg:pipeline}
\renewcommand{\algorithmicrequire}{\textbf{Input:}}
\renewcommand{\algorithmicensure}{\textbf{Output:}}
\begin{algorithmic}[1]
\REQUIRE For each sample $i=1, \dots, n$, the discrete  measurements $\{y^{\mathrm{obs}}_{i,j}(t_{i\ell})\}$ for each node $j \in V$ and the covariates $\mathbf{x}_i\in\mathbb{R}^q$.
\STATE{\textbf{Data Processing.} Continuous covariates are standardized (centered and scaled to unit variance); categorical covariates are encoded according to (\ref{eq:one_hot_enc}).}
\STATE{\textbf{Smoothing.} For each sample $i$ and node $j$, smooth the discrete observations to obtain a functional curve $\widetilde y_{i,j}\in\mathcal{H}$.} \label{step:smoothing}
\STATE{\textbf{Node-wise neighbourhood estimation.} For each $j \in V$ do:
\begin{itemize}
\item[(a)] \textbf{Finite dimensional approximation to vov regression} 
\begin{itemize}
\item[(a.1)] Define an orthonormal basis and retain the first $M$ elements.
\item[(a.2)]For each node $k \in V$ and sample $i$, compute projection score
$\mathbf{a}^{(j)}_{i,k}$. \label{step:dim_red}
\end{itemize}
\item[(b)] \textbf{Penalized coefficient estimation} with optimal hyperparameters selected via SCV (Appendix \ref{sec:hyperparameters}):
\begin{itemize}
\item[(b.1)] Estimate the coefficient matrices $\{\widehat{\mathbf{B}}^{c}_{k}\}$ for $k \in V \setminus \{j \}$ and $c=0, \dots, q$ by solving the group-lasso penalized least square problem described by \eqref{eq:optimization_prob}. 
\item[(b.2)] For each $c=0,\ldots,q$ form the set of neighbours $\widehat{\mathcal{N}}_j^c$ as described in (\ref{eq:N_def_thr})
\end{itemize}
\item[(c)]\textbf{Inference of the direction of change} For each $k\in\widehat{\mathcal{N}}_j^c$, compute the variation rate $VR^c_{kj}$ as in \eqref{eq:relative_eff}.
\end{itemize}}
\STATE{\textbf{Graph construction.} For each $c=0,\ldots,q$, aggregate $\{\widehat{\mathcal{N}}_j^c\}_{j=1}^p$ and obtain an undirected edge set $\widehat{E}^c$ via OR- or AND-symmetrization (\eqref{eq:or_symm}-\eqref{eq:and_symm}), yielding $\widehat{G}_c=(V,\widehat{E}^c)$.}
\STATE{\textbf{Edge weighting.} For each $(k,j)\in\widehat{E}^c$, define the weight $w_{kj}^c$ as in \eqref{eq:edge_weights} to discriminate attenuated vs strengthened dependendences.}
\ENSURE{Estimated graphs $\{\widehat{G}_c\}_{c=0}^q$ and edge weights $\{w_{kj}^c\}_{c=1}^q$.}
\end{algorithmic}
\end{algorithm}

\section{Simulation study}\label{sec:results_simulation}

We tested the finite-sample performance of the proposed approach using simulated data. In the following, the dataset simulation and the model's performance are presented.

\subsection{Data generation}\label{sec:data_gen}
To be able to quantitatively compare our results with those of existing methods directly estimating the differential network between two groups \citep{zhao2022}, we model the presence of a single binary external variable representing a grouping factor that splits the sampled data into two groups: $Gr^0$, taken as the reference group, and  $Gr^1$. Hence, in our simulation $q=1$ and $x_{i,1} = 0$ for all the observations $i$ that belong to the reference group $Gr^0$, while $x_{i,1} = 1$ for all the other observations within $Gr^1$.
Appendix~\ref{sec:add_sim_analysis_res}
presents additional simulations with $q=2$ covariates, including a binary grouping factor and a continuous variable, to demonstrate the model's flexibility in handling multiple heterogeneous external variables simultaneously.

We assume the functional data are observed at $\tau=100$ time points equally spaced in (0, 1]. For all $i = 1, \dots, n$ and all $j = 1, \dots, p$, the observed values are generated as 
\begin{displaymath}
y_{i,j}^{obs}(t_k) = \boldsymbol{\alpha}_{i,j}^{\top} \mathbf{f}(t_k) + \eta_{i,j}(t_k) \quad \quad k = 1, \dots, \tau
\end{displaymath}
where $\mathbf{f}(t_k) = \left(f_1(t_k), \dots, f_{M^*}(t_k) \right)^{T}$ collects the value of the first $M^*$ Fourier basis  functions \citep{ramsay2005} at the $k$-th observed time points; in our simulations we set $M^* = 15$; $\eta_{i,j}(t_k)$ is drawn from a zero-mean Gaussian distribution with variance $\sigma^2 = 0.5$; $\boldsymbol{\alpha}_{i,j}^{\top} = (\alpha_{i,j1}, \dots, \alpha_{i,jM^*}) \in \mathbb{R}^{M^*}$ are scoring factors such that for each observation, $i=1, \dots, n$, the vector $(\boldsymbol{\alpha}_{i,1}^{\top}, \dots, \boldsymbol{\alpha}_{i,p}^{\top})^{\top} \in \mathbb{R}^{pM^*}$ is randomly drawn from a multivariate Gaussian distribution with zero mean and covariance matrix $\mathbf{\Sigma}_i = (\mathbf{\Theta}^0 + x_{i,1} \mathbf{\Theta}^1)^{-1}$. 
Here, $\mathbf{\Theta}^0$ and $\mathbf{\Theta}^1$ are symmetric matrices of size $pM^* \times pM^*$ encoding respectively the population network $G^0$, that in our simulations describes the conditional interdependencies within data from the reference group $Gr^0$, and the differential graph $G^1$, collecting pairs of nodes whose conditional interdependencies is different in the two observed groups. The precision matrix $\mathbf{\Theta}^0 + \mathbf{\Theta}^1$ describes the conditional interdependencies within data from group $Gr^1$.

\begin{figure}
\centering{
\includegraphics[width=\textwidth]{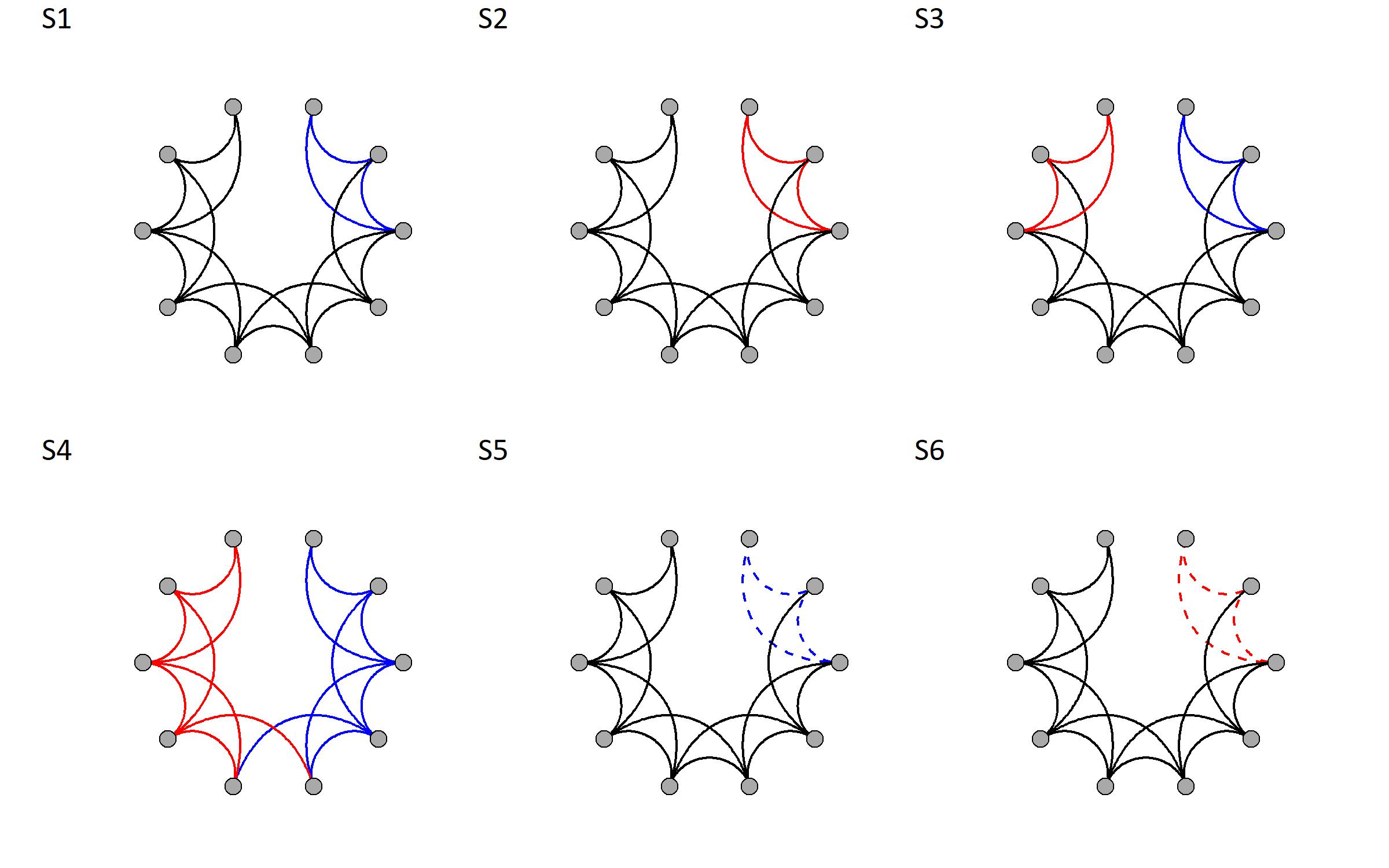}}\caption{Visual illustration of the graphical models involved in the simulated scenarios S1-S6 described in detail in Section \ref{sec:data_gen}. In all graphs: black solid lines depict pairwise conditional interdependences that are common to the two groups, $Gr^0$ and $Gr^1$; red and blue solid lines depict pairwise conditional interdependencies that are present only in one group, the reference group and the other group, respectively; red and blue dashed lines represent interdependencies that are common to the two groups but that are stronger in the reference or in the other group, respectively.}\label{fig:graph_sims}
\end{figure}

The precision matrices $\mathbf{\Theta}^0$ and $\mathbf{\Theta}^0 + \mathbf{\Theta}^1$ are defined by modifying the block-banded matrix $\mathbf{\Theta} \in \mathbb{R}^{pM^* \times pM^*}$ whose non-zero blocks are given as 
\begin{equation}
\label{eq:theta_structure}
\mathbf{\Theta}_{kj} =
\left\{
\begin{array}{cl}
\mathbf{T} & \text{if } k=j \\[4pt]
0.4\,\mathbf{A} & \text{if } k \in \{j-1, j+1\} \\[4pt]
0.2\,\mathbf{I}_{M^*} & \text{if } k \in \{j-2, j+2\}
\end{array}
\right.
\end{equation}
where $\mathbf{T} \in \mathbb{R}^{M^* \times M^*}$ is a Toeplitz matrix such that $T_{vv} = 1$ and $T_{vw} = 0.5 |v-w|$ for all $v\neq w$; $\mathbf{A} \in \mathbb{R}^{M^* \times M^*}$ is a tridiagonal matrix such that $A_{vv}=1$ and $A_{v(v+1)} = A_{(v+1)v} =0.5$; $\mathbf{I}_{M^*} \in \mathbb{R}^{M^* \times M^*}$ is the identity matrix. In detail, we initially set both the precision matrices equal to $\mathbf{\Theta}$ and then we modified them in order to simulate the following scenarios, denoted as S1-S6 and schematically depicted in Fig.~\ref{fig:graph_sims}.
\begin{itemize}
\item[\textit{S1.}] The samples within $Gr^0$ show fewer pairwise conditionally interdependencies than those in $Gr^1$. Towards this end we set $\mathbf{\Theta}^0_{kj}=\mathbf{0}$ for all $j,k \in \{1, \dots, \lfloor \frac{p}{3} \rfloor \}$.
\item[\textit{S2.}] The samples within $Gr^0$ show some additional interdependencies with respect to those in $Gr^1$. To simulate this condition we set $(\mathbf{\Theta}^0 + \mathbf{\Theta}^1)_{kj} = \mathbf{0}$ for all $k,j \in  \{1, \dots, \lfloor \frac{p}{3} \rfloor \}$.
\item[\textit{S3.}] $Gr^0$ and $Gr^1$ involve two different but intersecting sets of edges. This condition is simulated by setting $\mathbf{\Theta}^0_{kj}=\mathbf{0}$ for all $j,k \in \{1, \dots, \lfloor \frac{p}{3} \rfloor \}$ and $(\mathbf{\Theta}^0 + \mathbf{\Theta}^1)_{kj} = \mathbf{0}$ for all $k,j \in  \{p-\lfloor \frac{p}{3} \rfloor+1, \dots, p \}$
\item[\textit{S4.}] $Gr^0$ and $Gr^1$ involve two non-intersecting sets of edges. In this case we $\mathbf{\Theta}^0_{kj}=\mathbf{0}$ for all $j,k \in \{1, \dots, \lfloor \frac{p}{2} \rfloor + 1  \}$ and $(\mathbf{\Theta}^0 + \mathbf{\Theta}^1)_{kj} = \mathbf{0}$ for all $k,j \in  \{p-\lfloor \frac{p}{2} \rfloor, \dots, p \}$
\item[\textit{S5.}] $Gr^0$ and $Gr^1$ show the same pairwise conditional interdependencies, but some of the relationships are stronger in $Gr^1$. To simulate this scenario for all $j,k \in \{1, \dots, \lfloor \frac{p}{3} \rfloor \}$, $j \neq k$, we set $\mathbf{\Theta}^0_{kj}=0.25\mathbf{\Theta}_{kj}$.
\item[\textit{S6.}] Similarly to the previous scenario, $Gr^0$ and $Gr^1$ show the same pairwise conditional interdependencies, but in $Gr^1$ some of the relationships are weaker. To simulate this scenario we set $(\mathbf{\Theta}^0 + \mathbf{\Theta}^1)_{kj}= 0.25 \mathbf{\Theta}_{kj}$ for all $j,k \in \{1, \dots, \lfloor \frac{p}{3} \rfloor \}$, $j\neq k$.

\end{itemize}

\subsection{Benchmark method and evaluation metrics}
\label{subsec:benchmark_and_eval_metrics}
In simulation studies, we compare the proposed estimator with FuDGE and three joint graphical-lasso baselines (FGL, FFGL, FFGL2) introduced by \citet{zhao2022}.
All benchmark methods operate on a finite-dimensional representation of the functional observations (e.g., FPCA scores), and performance is assessed in terms of recovery of the differential graph $G^1$.

\paragraph{Benchmark methods.}
\emph{FuDGE} (Functional Differential Graph Estimation) is a direct estimator of the functional differential graph. Rather than estimating the two group-specific graphs and subtracting them, FuDGE targets the difference of the (finite-dimensional) precision operators, $\mathbf{\Theta}^1$, through a convex quadratic loss constructed from the sample covariance matrices of the score vectors, combined with a penalty that promotes sparsity of $\mathbf{\Theta}^1$ at the block (edge) level. As a result, FuDGE outputs only the differential network, which can be appealing when the two underlying graphs are dense but their difference is sparse.

\emph{FGL} (Fused Graphical Lasso) is a joint estimation approach originally developed for multiple related Gaussian graphical models. It estimates the precision matrices of the two groups simultaneously using an $\ell_1$ penalty to induce sparsity within each group and an additional fused $\ell_1$ penalty on entry-wise differences to encourage similarity of edge values across groups. In the functional setting, FGL is applied directly to the (vectorized) score representation and does not explicitly exploit the natural block structure induced by the basis expansion.

\emph{FFGL} and \emph{FFGL2} are functional extensions of FGL that incorporate block structure. Both methods jointly estimate $\mathbf{\Theta}^0$ and $\mathbf{\Theta}^0+ \mathbf{\Theta}^1$ with a block-wise sparsity penalty (Frobenius norms of off-diagonal blocks). They differ in how similarity across groups is enforced: FFGL uses a fused Frobenius penalty on differences between corresponding blocks, whereas FFGL2 retains an element-wise fused $\ell_1$ penalty within blocks. Differential edges are then obtained by differencing the two estimated precision matrices.

\paragraph{Relation to the proposed method.}
Unlike FuDGE, which estimates only the differential graph $\mathbf{\Theta}^1$ and FGL and FFGL that focus their estimation on the graphs associated to the two groups $\mathbf{\Theta}^0$ and $\mathbf{\Theta}^0 + \mathbf{\Theta}^1$, our approach is formulated to estimate simultaneously the group component $\mathbf{\Theta}^0$ and the differential graph  $\mathbf{\Theta}^1$ within a single optimisation problem. This joint parameterisation allows us to recover both group-specific networks $Gr^0$ described by $\mathbf{\Theta}^0$ and $Gr^1$ given by $$\widehat{\mathcal{N}}_j^{Gr^1} = \left\{k \in V \setminus \{j\} : ||\widehat{\mathbf{B}}_k^0 + \widehat{\mathbf{B}}_k^1||_{F} > \varepsilon_j \right\} \,$$  while still enabling a fair comparison on the differential-network recovery task used in the simulations.

\paragraph{Evaluation metrics.}
Across simulations, we evaluate recovery of the target differential adjacency structure using precision, true positive rate (TPR), false positive rate (FPR), and the $F_1$ score on the undirected graph obtained after symmetrization. Specifically, we report
$\mathrm{Precision}=\mathrm{TP}/(\mathrm{TP}+\mathrm{FP})$, $\mathrm{TPR}=\mathrm{TP}/(\mathrm{TP}+\mathrm{FN})$,
$\mathrm{FPR}=\mathrm{FP}/(\mathrm{FP}+\mathrm{TN})$, and
$F_1 = 2\,\mathrm{TP}/(2\,\mathrm{TP}+\mathrm{FP}+\mathrm{FN})$.
Metrics are computed on the upper-triangular part of the symmetric adjacency matrices. Results are primarily summarized in terms of $F_1$, as it provides a single measure that balances precision and recall.

For FuDGE and the related baselines, the public simulation code release does not include an automatic, data-driven hyperparameter selection routine. Therefore, for these methods, we computed the metrics over the full hyperparameter sequences provided by the respective implementations and, for each simulation setting, reported the best-performing configuration (highest $F_1$) in Fig.~\ref{fig:f1_comparison_fudge}. This choice yields an optimistic estimate of benchmark performance relative to our approach, which selects tuning parameters in a fully data-driven manner.

\subsection{Results}
To test the impact of sample size and number of nodes on our proposed method, we focused on scenario $S3$, and we generated 10 simulated data for each combination of $p \in \{10, 15, 25, 50\}$ and $n \in 2 \cdot \{50, 75, 100, 150, 200\}$ as described in Section \ref{sec:data_gen}. The two groups, $Gr^0$ and $Gr^1$, have equal sample sizes. 

\begin{figure}
\centering
\includegraphics[width=\textwidth]{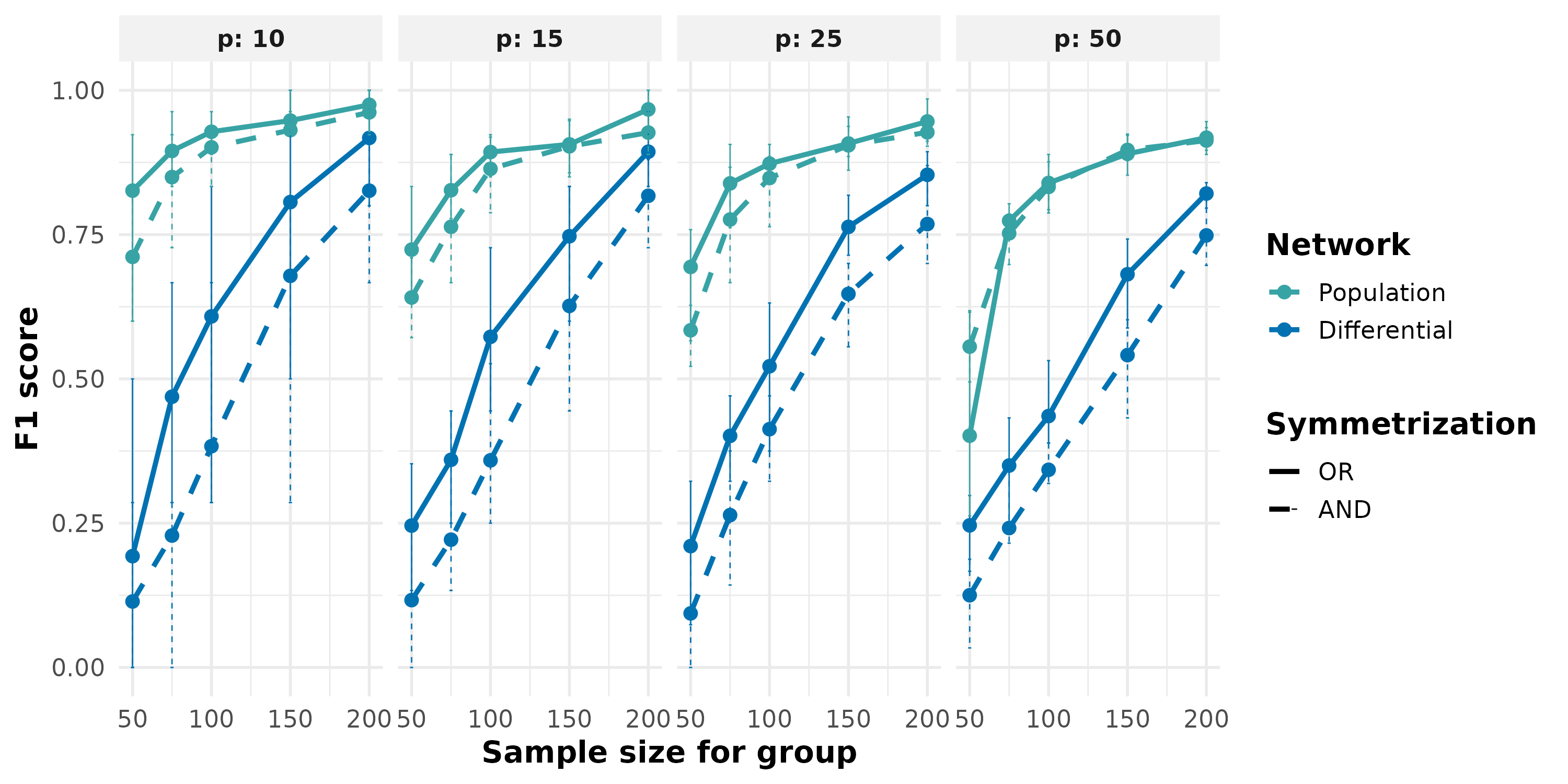}\caption{Performance in estimating the population network $G^0$ and the differential graph $G^1$ for different numbers of nodes $p$, sample sizes $n$, and symmetrization
strategies. For each combination of $p$ and $n$ the plots show minimum, mean and maximum F1 score reached over 10 simulated data generated according to scenario S3.}\label{fig:F1score_nostro}
\end{figure}

Fig. \ref{fig:F1score_nostro} shows that the \texttt{OR} symmetrization strategy in Eq. (\ref{eq:or_symm}) should be preferred over the \texttt{AND} strategy in Eq. (\ref{eq:and_symm}): while the two strategies provide comparable results for the population network $G^0$, for the differential network $G^1$ the F1 score of the \texttt{OR} strategy is systematically higher. For this reason, in the remainder of the section, only results for this strategy are shown.
Furthermore, Fig. \ref{fig:F1score_nostro} shows that the estimates of the population network are usually more accurate than the estimates of the differential network. However, with enough samples, our approach provides accurate estimates of both networks: when the group sample size, $n/2$, is greater than 150 ($n/2 \geq 200$ for $p=50$), the average F1 score exceeds 0.75 for the differential network and approaches 0.9 for the population network. Notably, as shown in Appendix \ref{subsubsec:sample_size_req} Fig. \ref{fig:FPRscore_1_cov_setting}, the FPR remains close to zero for all the considered sample sizes, meaning that although for the lowest sample sizes the support of the networks cannot be recovered, the estimated pairwise interdependencies result to be correct.

\begin{figure}
\centering
\includegraphics[width=\textwidth]{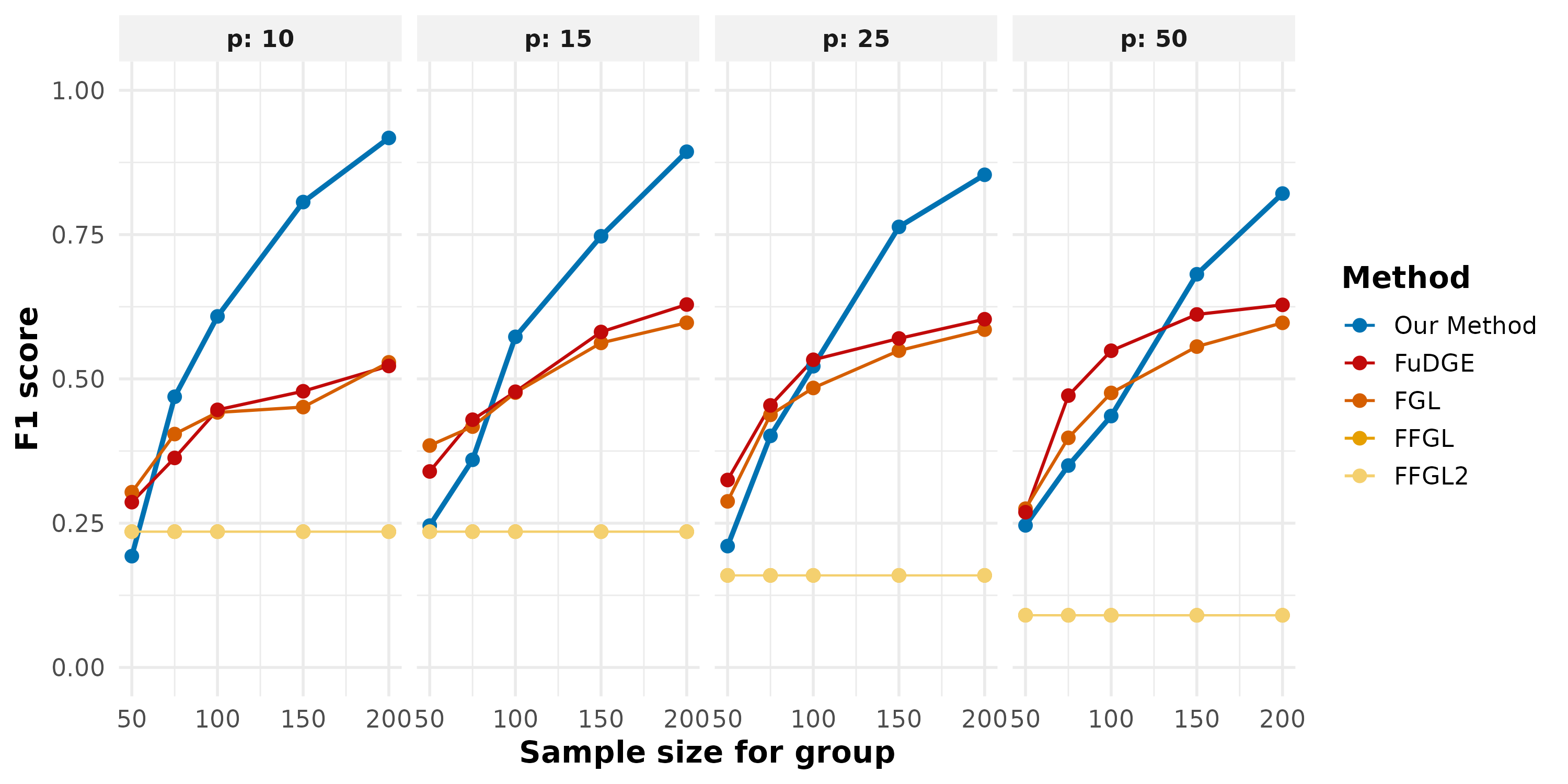}\caption{Comparison of the performance in estimating the differential network $G^1$ of our approach with respect to state-of-the-art methods. Plots show the average F1 score over the 10 simulated datasets considered in Fig. \ref{fig:F1score_nostro}. Note that the F1-scores for FFGL coincide with those of FFGL2.}\label{fig:f1_comparison_fudge}
\end{figure}

\begin{figure}
\centering
\includegraphics[width=\textwidth]{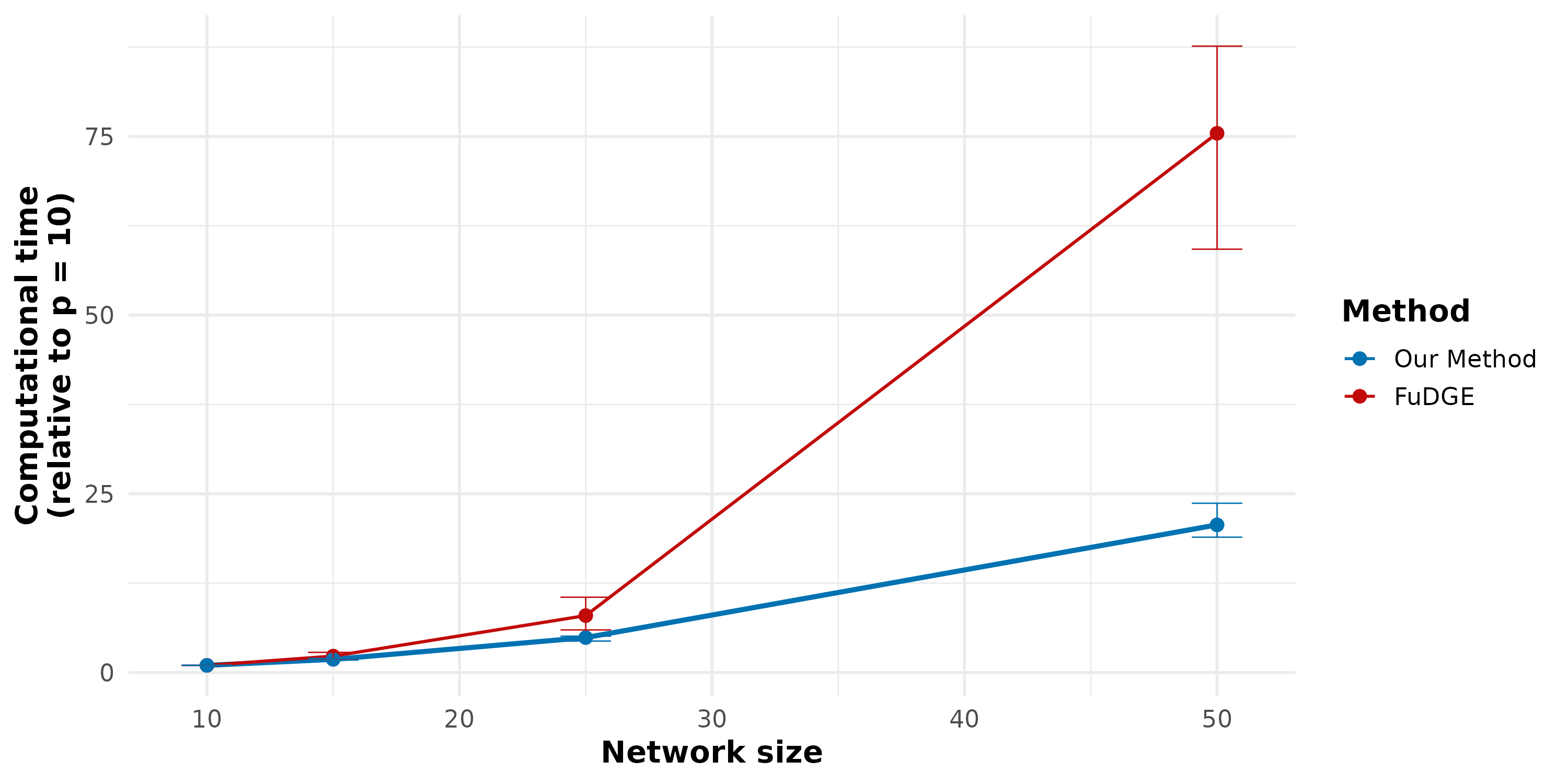}\caption{Computational scalability of the proposed method and of FuDGE as network size increases. For each network size, $p$, and a fixed sample size of 200 observations per group, runtime was normalized by the average computational time of the corresponding method at $p=10$. Points represent the mean relative runtime across 10 simulated datasets, while error bars denote the minimum and maximum values observed across replicates.
}\label{fig:comp_time_comparison}
\end{figure}

As a touchstone, Fig. \ref{fig:f1_comparison_fudge} compares the proposed approach with four competitors. Among the latter, the best performance was achieved by FuDGE, whose F1 scores associated with the differential network $G^1$ remained below 0.65 even at the highest tested sample sizes. Our approach also demonstrates superior computational scalability compared with FuDGE, as shown in Fig.~\ref{fig:comp_time_comparison}. While FuDGE operates on the full precision matrices associated with the functional processes, our method relies on a neighbourhood-selection strategy that can be fully parallelized across nodes. To compare scalability independently of baseline implementation differences, runtimes were normalized by the average computational time of each method at $p=10$. Although the two methods exhibit comparable runtimes for small networks, their computational costs diverge substantially as network size increases. At $p=50$, FuDGE requires approximately 75 times its baseline runtime, whereas our approach requires less than 25 times its corresponding baseline.

\begin{figure}
\centering{
\includegraphics[width=\textwidth]{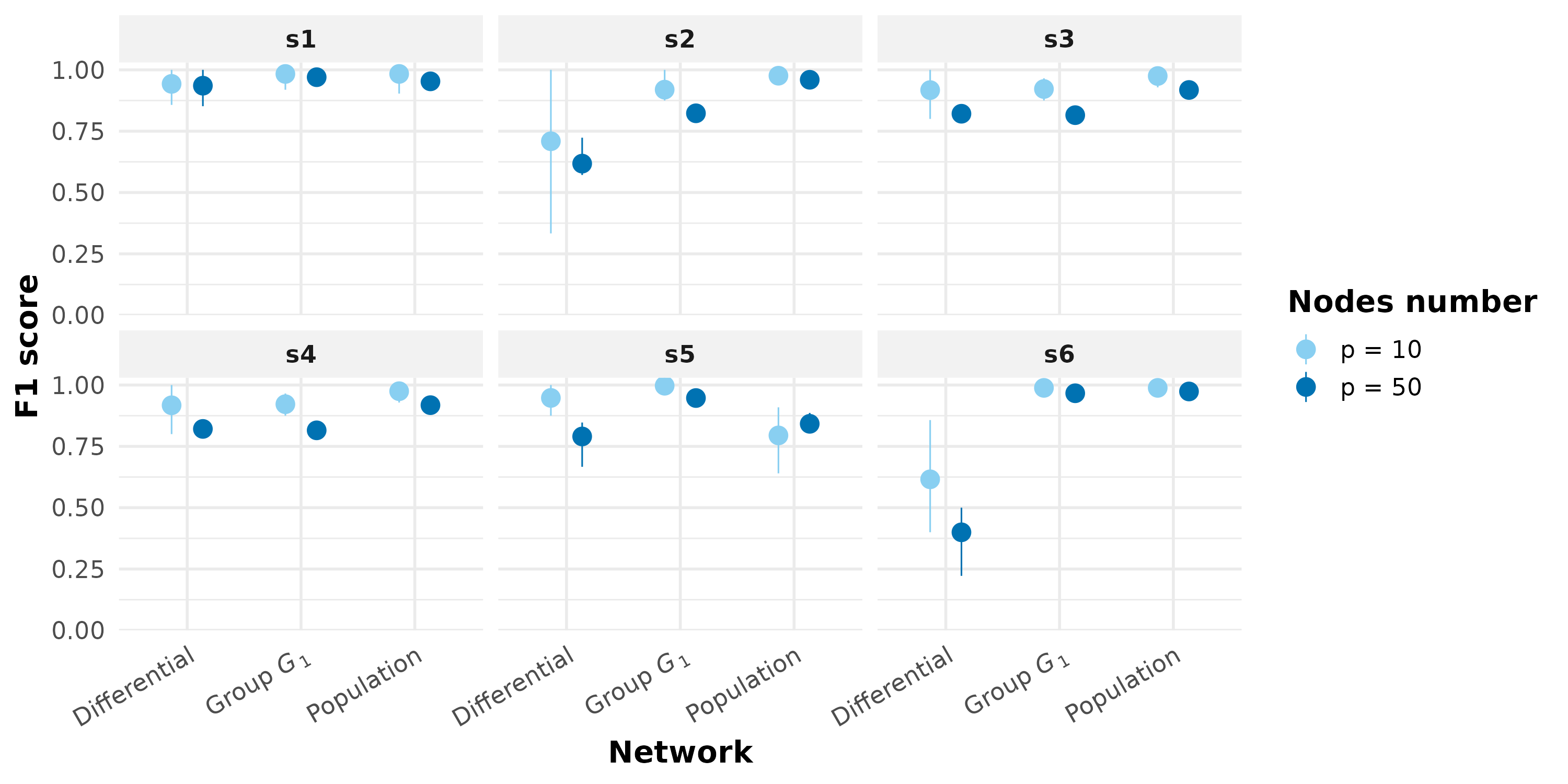}}\caption{Performance in estimating the differential network $G^1$ and the graphical models associated to the groups $Gr^0$ (population network) and $Gr^1$, under the scenarios in Fig. \ref{fig:graph_sims} when the number of nodes is $p \in \{10, 50\}$ and the sample size of each group is $n/2 = 200$. For each scenario, the plots show the minimum, mean, and maximum F1 scores across 10 simulated datasets.}\label{fig:f1_diff_scenarios}
\end{figure}

As a further experiment, we tested the robustness of our results when varying the underlying graphical structure. Towards this end, we fixed $p \in \{10, 50 \}$ and the group sample size to $n/2 = 200$ and we generated 10 simulated data for each scenario depicted in Fig. \ref{fig:graph_sims}. As can be seen from Fig. \ref{fig:f1_diff_scenarios}, our approach reached satisfying results in all the considered scenarios. In scenarios S1, S3, S4, and S5, the averaged F1-score for the differential network was close to 1 for $p=10$ and to 0.8 for $p=50$. Slightly worse results were obtained in scenarios S2 and S6, where the differential network includes only pairwise interdependencies that are stronger in the reference group $Gr^0$. Note that, in these scenarios, the support of the graphical models of the two groups $Gr^0$ and $Gr^1$ is correctly identified, suggesting that this behavior may be partially due to the penalization term of the loss function in (\ref{eq:pen_least_square}), which forces sparsity in the estimates of the differential network.

\section{Case study: effect of alcohol abuse on EEG functional connectivity}\label{sec:results_experdata}

We apply our method to a publicly available EEG dataset \citep{zhang1995,ingber1997} involving $ n = 122$ participants, including $77$ individuals diagnosed with alcohol use disorder (AUD) and $45$ control subjects. EEG recordings were collected during an object recognition task using a standard $61$-sensor EEG cap, with one ground sensor placed on the nose and two bipolar derivations recording the vertical and horizontal electrooculogram (EOG). To obtain results comparable with those presented by \cite{zhao2022} we thus fixed $p=64$ and we relied on their preprocessed EEG data, where signals were filtered within the $\alpha$-frequency band (8–12.5 Hz)  \citep{Knyazev2007, Zhu2014}.

We applied our proposed method equipped with the \texttt{OR} symmetrization strategy, treating the binary covariate that distinguishes the controls (chosen as the reference group $Gr^0$) from the group $Gr^1$ of individuals with AUD as an external variable.

\begin{figure}
    \centering
    \includegraphics[width=\linewidth]{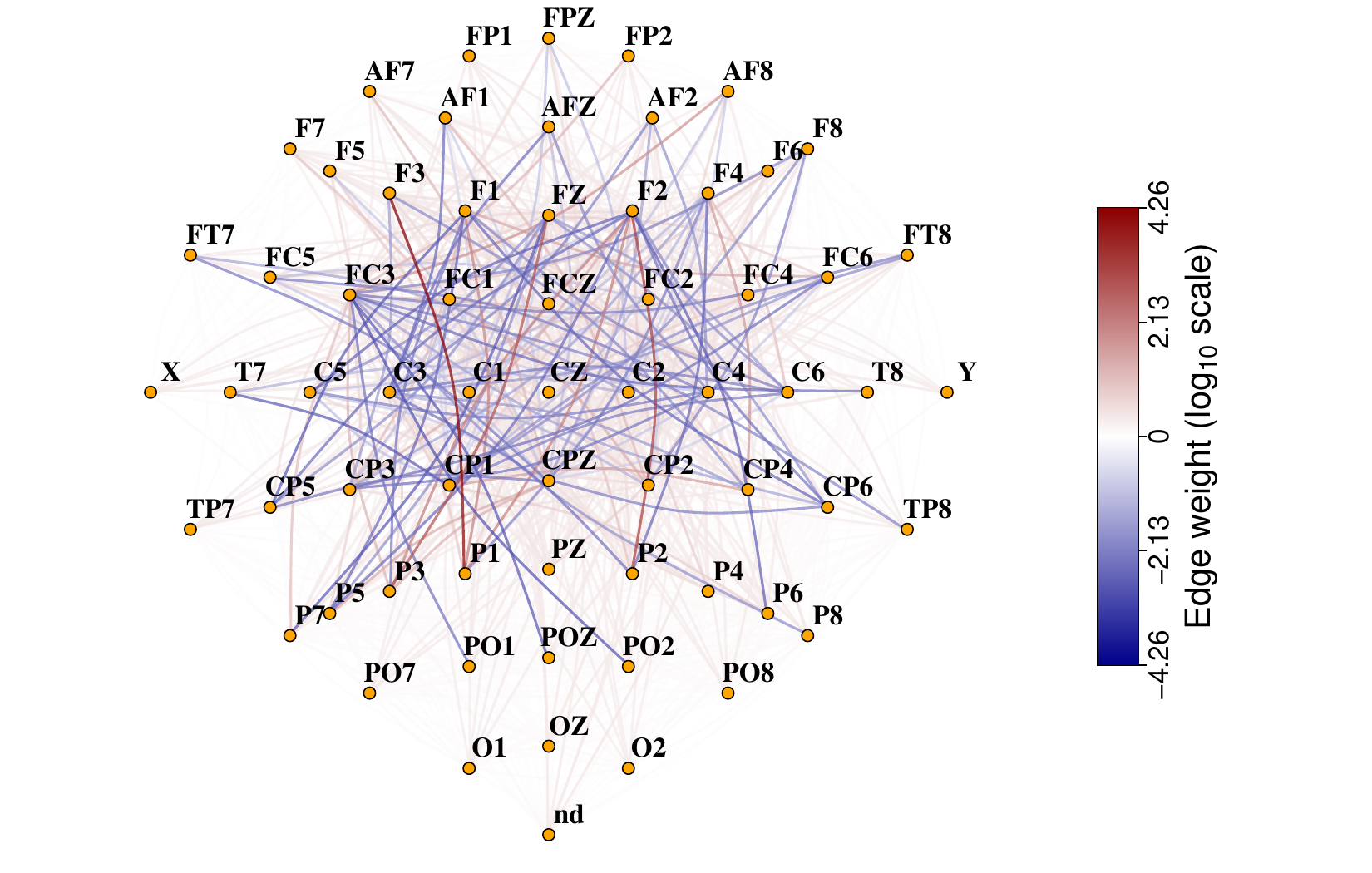}
    \caption{Differential network characterizing changes in EEG functional connectivity in individuals with AUD with respect to controls. Edges are coloured based on the weights defined in Eq.(\ref{eq:edge_weights}) so that blue and red edges indicate a reduction and an increase in the interaction strength, respectively.}\label{fig:res_EEG_data}
\end{figure}

Appendix \ref{sec:whole_output_EEG} summarizes the output provided by our approach, while Fig. \ref{fig:res_EEG_data} focuses on the estimated differential network describing changes in EEG functional connectivity related to AUD. In this figure, strengthening and attenuation of the interaction strength are discerned through the decimal logarithm of the weights in Eq. (\ref{eq:edge_weights}).

Our results mainly indicated a decreased functional connectivity among the fronto-central areas and between the frontal areas and the posterior-occipital regions. This result is consistent with previous studies reporting impaired connectivity in individuals with alcohol dependence \citep{deBruin2005,herrera2016}, although most previous works focused on a different recording condition where subjects are resting. Furthermore, the regions involved in our differential network are similar to those identified by \cite{zhao2022} using FuDGE, with the main difference being that most regions in their network were connected to node X. It is worth noting that channel X was not included in the analysis of the original study by \cite{zhang1995}, as it is an unconventional channel for the standard EEG cap used in their work and may have served as a sensor for controlling EOG artifacts rather than for measuring brain activity. More importantly, while we recognize a reduction in connection strength, FuDGE provides no information on the direction of the change. As a result, our framework provides a more interpretable characterization of connectivity alterations and facilitates the investigation of clinically relevant questions such as those considered in this application.

\section{Discussion}\label{sec:discussion}
Motivated by the problem of estimating group-specific EEG functional connectivity, we proposed a neighborhood selection procedure based on functional-on-functional regression for conditional Gaussian functional graphical models with an arbitrary number of continuous and discrete covariates. Given EEG recordings from $n$ individuals, the resulting pipeline returns (i) a population-level graph $G^0$ capturing conditional dependencies shared across subjects, and (ii) a covariate-specific differential graph $G^c$ for each covariate, highlighting the connections whose strength varies with $X_c$.

The proposed framework contributes to the literature in two main ways. First, it extends neighborhood selection from unconditional functional graphical models to a conditional setting, providing a unified, practical pipeline for estimating, within a single model, both the baseline population network and the covariate-specific differential components. This joint decomposition enables two complementary types of inference: it allows one to reconstruct group-specific connectivity patterns by combining baseline and covariate terms, while simultaneously isolating differential networks to support clinical insights. Second, by leveraging the estimated regression coefficients, our method not only identifies which pairwise interdependencies vary with each covariate, but it also characterizes whether a variation in the covariate is associated with an increase or decrease in the interaction strength.
This information is not readily available in most current methods, which focus solely on differential edge support.

Our simulation studies and real-data application support the practical advantages of the method. From a methodological standpoint, we introduced a computationally efficient estimator based on a block-structured group-lasso objective solved by an ADMM routine, combined with a data-driven SCV procedure for node-specific hyperparameter selection, including both the regularization level controlling sparsity in the regression coefficients and the threshold used to identify non-negligible edges. Empirically, across a broad set of simulation scenarios, the method achieved competitive or superior edge-recovery performance relative to existing approaches, particularly as the number of subjects increased. In addition, by exploiting that neighbor selection can be performed independently for each node, our approach is parallelized across nodes, resulting in an overall computational cost that scales favorably in high-dimensional settings, including values of $p$ comparable to commonly used EEG montages. In contrast, methods such as FuDGE, which directly estimate the full precision matrix, offer fewer opportunities for parallelisation and exhibit less favourable scalability as network dimensionality increases. This makes our approach better suited to real-world applications where the number of sensors can be large. Our method remained robust across the considered data-generating settings, and the real-data analysis produced connectivity patterns consistent with prior neurophysiological findings, supporting the reliability of the inferred networks. Although motivated by EEG connectivity analysis, the proposed framework is applicable to a broad range of multivariate functional data settings, including domains such as functional MRI, where increasingly large datasets are becoming available~\citep{bellec2017neuro}.

A key limitation emerges in the low-sample regime, where reconstructing differential networks is more challenging. Specifically, in low sample size settings, such as those commonly encountered in clinical EEG studies, the proposed approach resulted rather conservative. As shown by the analysis of the $F_1$ score (Fig. \ref{fig:F1score_nostro}) and the FPR (Fig. \ref{fig:FPRscore_1_cov_setting}) in these settings, our approach fails to recover the full support of the differential network, but the estimated differential interdependencies are unlikely to be spurious. This behavior is expected given the high dimensionality of the parameter space induced by the functional representation and by the presence of multiple covariate-specific coefficient blocks. More generally, sample size requirements depend on both the number of covariates included in the model and the nature and magnitude of the covariate-related variation. Among the scenarios in Fig.~\ref{fig:f1_diff_scenarios}, slightly lower $F_1$ scores are observed in scenarios S2 and S6, where the reference group $Gr^0$ presents the densest conditional interdependency structure. In both cases, the model must estimate the full population network associated with $Gr^0$ while simultaneously identifying the
regression parameters encoding the covariate-related reduction in conditional interdependencies in $Gr^1$, increasing the effective number of non-zero parameters to recover. Moreover, the current penalization is primarily block-wise, encouraging sparsity at the edge level but not within-edge structure. As a consequence, when sample sizes are limited, the estimator may either retain or discard an entire edge block, even when only a subset of within-block coefficients is reliably supported by the data.
Given these considerations, to improve power in low-dimensional setting we advise: (i) carefully selecting which covariates to include in the model; (ii) when dealing with categorical covariates representing grouping factor, designating as the reference group the one expected to present lower conditional interdependency strength; (iii) reducing the number of nodes $p$ by aggregating brain regions into larger region of interest.

Furthermore, several directions are promising for improving performance in such settings. A natural extension is to incorporate additional penalties that act within blocks (e.g., sparse-group penalties), enabling the model to reduce the effective number of parameters without forcing the complete removal of an edge. A second direction concerns scalability with respect to the number of covariates. While the current framework accommodates an arbitrary set of covariates, inference may become sample-limited as $q$ grows. To address this, future work will consider introducing an additional layer of regularization at the covariate level (e.g., a penalty on entire covariate-specific coefficient groups), yielding a data-driven mechanism for identifying influential covariates while shrinking negligible covariate effects towards zero. This extension would support applications in which many clinical or behavioral variables are available, while maintaining stable network estimation under realistic sample sizes.

\begin{acknowledgement}
The present research has been partially supported by MUR, grant Dipartimento di Eccellenza 2023-2027 awarded to the Department of Mathematics of Politecnico di Milano (A.M. and F.I.) and to the Department of Mathematics of The Università di Genova (L.C. and S.S.). \\
S.S. acknowledges the support of the PRIN PNRR 2022 Project 'Computational mEthods for  Medical Imaging (CEMI)' 2022FHCNY3, cup: D53D23005830006.
\end{acknowledgement}
\vspace*{1pc}

\noindent {\bf{Data Availability}}

The EEG data that support the findings of this study are openly available in the GitHub repository associated to the work by \cite{zhao2022} available at https://github.com/boxinz17/FuDGE/tree/master/EEG/. 

\vspace*{1pc}

\noindent {\bf{Code Availability}}
All code required to reproduce the analyses and figures in this manuscript is available at: \texttt{https://github.com/AlessiaMapelli/condFGM}.

\vspace*{1pc}
\noindent {\bf{Conflict of Interest}}

\noindent {The authors have declared no conflict of interest.}

\section*{Appendix}
\setcounter{subsection}{0}
\renewcommand{\thesubsection}{A.\arabic{subsection}}

\subsection{Optimization algorithm details}
\label{supp:alg_details}
The developed optimization algorithm builds on \cite{zhao2024} and extends their Alternating Direction Method of Multipliers (ADMM, \cite{Boyd2011}) routine for solving the optimization problem in (\ref{eq:pen_least_square}) to accommodate a user-specified number of covariates. The resulting implementation is designed to scale from sequential execution to parallel runs in high-performance computing (HPC) environments. Specifically,
Eq. (\ref{eq:pen_least_square}) can be reformulated in a more compact way as 
\begin{equation}\label{eq:pen_least_square_matrix}
\widehat{\boldmcal{B}} \in \argmin_{\boldmcal{B}\in \mathbb{R}^{M \times M (p-1)(q+1)}} \left\{ \frac{1}{2n} \left|\left| \mathbf{A}_p - (\boldmcal{A}*\boldmcal{X}) \boldmcal{B}  \right|\right|_F^2 + \lambda_p \sum_{k=1}^{(p-1)(q+1)} ||\boldmcal{B}_k||_F\right\} \, .
\end{equation}
In Eq. (\ref{eq:pen_least_square_matrix}), $\boldmcal{B} = [\mathbf{B}_1^0, \dots, \mathbf{B}_{p-1}^0, \dots, \mathbf{B}_1^q, \dots, \mathbf{B}_{p-1}^q] \in \mathbb{R}^{M \times M (p-1) (q+1)}$ and $\boldmcal{B}_k$ scans the $M \times M$ coefficient matrices within $\boldmcal{B}$; $\boldmcal{A} = \left[\mathbf{A}_1, \mathbf{A}_2, \dots, \mathbf{A}_{p-1} \right] \in \mathbb{R}^{n \times M(p-1)}$, where for all $j=1, \dots, p$, 
$$\mathbf{A}_j = \left[ \begin{array}{c}
\mathbf{a}_{1,j}^{\top} \\
\vdots \\
\mathbf{a}_{n,j}^{\top}
\end{array} \right] \in \mathbb{R}^{n \times M} \, ;$$ 
$\boldmcal{A}*\boldmcal{X} \in \mathbb{R}^{n \times M (p-1) (q+1)}$ denotes the matrix obtained by multiplying element-wise each column of $\boldmcal{A}$ by each columns of the matrix 
$$
\boldmcal{X} = \left[ \begin{array}{cc}
1 & \mathbf{x}_1 \\
\vdots & \vdots \\
1 & \mathbf{x}_n
\end{array} \right] \in \mathbb{R}^{n \times (q+1)} \, .
$$

The optimization problem in (\ref{eq:pen_least_square_matrix}) can be further rewritten as 
\begin{equation} \label{eq:optimization_prob}
\begin{gathered}
 \argmin_{\boldmcal{P}, \boldmcal{Q} \in \mathbb{R}^{M \times M(p-1)(q+1)}}  \left\{ \frac{1}{2n} \left|\left| \mathbf{A}_p - (\boldmcal{A}*\boldmcal{X}) \boldmcal{Q}  \right|\right|_F^2 + \lambda_p \sum_{k=1}^{(p-1)(q+1)} ||\boldmcal{P}_k||_F \right\}  \\
  \text{subject to } \boldmcal{P} = \boldmcal{Q}
\end{gathered}
\end{equation}
which can be solved via the Alternating Direction Method of Multipliers (ADMM, \cite{Boyd2011}).

\subsection{Choice of the basis}\label{sec:basis_def}
Both the smoothing step and the finite-dimensional approximation in (\ref{eq:serie_Y}) require specifying a basis for the Hilbert space $\mathcal{H}$.\\ In practice, the EEG signals are observed on finite grids and are affected by measurement error. For $i=1,\ldots,n$, $j \in V$, and observation times $\{t_{i\ell}\}_{\ell=1}^{L_i}\subset T$, we model the recorded values as
\[
y^{\mathrm{obs}}_{i,j}(t_{i\ell})=\widetilde y_{i,j}(t_{i\ell})+\eta_{i,j\ell},
\]
where $\widetilde y_{i,j}\in\mathcal{H}$ denotes the underlying smooth trajectory and $\{\eta_{i,j\ell}\}$ are i.i.d.\ noise terms. A standard approach is to approximate $\widetilde y_{i,j}$ through a basis expansion, $\widetilde y_{i,j}(t)\approx \sum_{r=1}^{R}\theta_{i,jr}\psi_r(t)$, and to estimate the coefficients $\theta_{i,jr}$ via (penalized) least squares, typically including a roughness penalty such as the integrated squared second derivative to encourage smoothness (\cite{ramsay2005functional}). This first-stage choice is driven by the qualitative features of the generating process: a Fourier basis is well suited for approximately periodic signals without localized bursts, a condition often encountered in resting-state EEG, whereas B-splines provide greater local adaptivity through their piecewise-polynomial construction and can better accommodate non-stationary patterns. The number of basis functions $R$ to include in this first step is a key tuning parameter, as it controls the flexibility of the reconstructed curves: larger $R$ allows the fit to capture increasingly complex temporal features, but may also lead to overfitting by reproducing measurement noise rather than the underlying signal. Prior knowledge should guide this choice. Within our applications, both in simulations and in the real-data analysis, we used a Fourier basis for the reconstruction step, as it naturally captures the oscillatory nature of EEG signals. We fixed the number of basis functions to $R=15$, which we found sufficiently large to provide adequate flexibility for representing complex trajectories while maintaining smoothness.

After smoothing, we construct the finite-dimensional representation used in the vector-on-vector model (\ref{eq:vov_regr_model}) by projecting each curve onto an orthonormal basis of $\mathcal{H}$. In particular, for each neighbourhood selection problem associated with a target node $j$, we adopt a basis $\{\phi^{(j)}_m\}_{m\geq 1}$ and compute the projection scores $a^{(j)}_{i,km}=\langle y_{i,k},\phi^{(j)}_m\rangle$, retaining the first $M$ components. Two main strategies are commonly employed for defining $\{\phi^{(j)}_m\}$: (i) a fixed, pre-specified basis (e.g., Fourier, B-spline, wavelet), chosen based on prior knowledge and computational convenience; (ii) a data-driven basis, most often functional principal components, obtained as the eigenfunctions of an estimated covariance operator. The FPCA basis yields the best $L_2$ approximation among all $M$-dimensional subspaces and therefore provides an efficient compression when $M$ is moderate. As noted by \cite{zhao2024}, using a data-dependent basis in place of a prior fixed one induces an additional error term in the regression model, reflecting uncertainty due to basis estimation.

Despite this additional source of variability, in the context of neighbourhood selection we favour data-driven bases over fixed choices, as a misspecified pre-selected basis may inadequately capture the dominant modes of variation and thereby bias edge selection. Accordingly, for each node $j$, we compute a node-specific basis $\{\phi^{(j)}_m\}$ intended to represent the variability of the signal at that node, and use it consistently across all regressions $k\neq j$. This choice aligns the score representations entering (\ref{eq:vov_regr_model}) and facilitates interpretation of the coefficient blocks. Our implementation supports FPCA-based dimensionality reduction. More generally, user-defined preprocessing can be incorporated whenever alternative bases or alternative FPCA constructions, as the target-driven FPCA basis for each neighborhood regression in the spirit of Zhao et al.\ (2024),  are of interest. The truncation level $M$ is again a key hyperparameter: it governs the fidelity of the finite-dimensional embedding (larger $M$ yields a more accurate representation of the underlying functions) but increases the dimension of the regression problem, and hence the computational burden and the sample size required for stable neighbourhood recovery. In FPCA-based implementations, we recommend selecting $M$ based on the proportion of explained variance, assessed across nodes to ensure a comparable level of functional information is retained throughout the network.

\subsection{Optimization hyperparameter selection}\label{sec:hyperparameters}

In the current implementation, two tuning parameters must be specified for each neighbourhood regression: the thresholding parameter $\epsilon_j$ in (\ref{eq:N_def_thr}) and the group-lasso penalty $\lambda_j$ in (\ref{eq:pen_least_square}). Both parameters are selected in a node-specific manner via selective cross-validation (SCV) following a similar procedure to \cite{zhao2024}, using an error criterion that combines predictive fit with a BIC-type complexity penalty.

To construct the candidate set for $\lambda$, we compute from the data a value $\lambda_{\max}$ large enough that, for every $j\in V$, the solution of (\ref{eq:pen_least_square}) is the all-zero estimator. We then consider a grid in log scale of penalties $\lambda\in(0,\lambda_{\max}]$ of user-specified size. Candidate values for $\epsilon$ are also user-defined. Given these sets, we select $(\lambda_j^\ast,\epsilon_j^\ast)$ by minimizing the average SCV criterion over $K$ folds, as follows.

We fixed node $p$, and for each candidate pair $(\lambda,\epsilon)$:
\begin{enumerate}
    \item Using the full data, estimate the block coefficient matrix
    $\widehat{\boldmcal{B}}(\lambda)$
    by solving (\ref{eq:optimization_prob}).
    
    \item  Define the estimated active block index set
    $\widehat{\mathcal{N}}_p(\lambda,\epsilon)$ by computing for each $c=0,...,q$\\ $\widehat{\mathcal{N}}_p^c(\lambda,\epsilon) = \left\{k \in V \setminus \{p\} : ||\widehat{\mathbf{B}}_k^c(\lambda)||_{F} > \varepsilon \right\} $
    \item For each fold $\nu=1,\ldots,K$, let $I_\nu$ denote the training indices and $I_{\mathrm{test},\nu}$ the corresponding test indices. Re-estimate the coefficients by unpenalized least squares on $I_\nu$, constrained to the screened set:

    \begin{displaymath}
    \begin{aligned}
    \tilde{\boldmcal{B}}^{(\nu)} \in \argmin_{\boldmcal{B}\in \mathbb{R}^{M \times M (p-1)(q+1)}} \left\{ \frac{1}{2n} \left|\left| \mathbf{A}_p - (\boldmcal{A}*\boldmcal{X}) \boldmcal{B}  \right|\right|_F^2 \right\}\\
    \text{s.t. } \tilde{\mathbf{B}}_k = 0 \text{  for all   } k \notin \widehat{\mathscr{N}}_p(\lambda,\epsilon).
    \end{aligned}
    \end{displaymath}
    
    \item For $i\in I_{\mathrm{test},\nu}$, define the refit residual 
    $ \widehat{\mathbf{re}}^{(\nu)}_i
    = \mathbf{A}_{i} -(\boldmcal{A}_{,i}*\boldmcal{X}_{,i}) \tilde{\boldmcal{B}}^{(\nu)}, $
    and compute the fold-specific SCV-RSS score
    \begin{displaymath}
    \mathrm{SCV\text{-}RSS}_\nu(\lambda,\epsilon)
    = \sum_{i\in I_{\mathrm{test},\nu}}\left\|\widehat{\mathbf{re}}^{(\nu)}_i\right\|_2^2
    + \log\!\bigl(|I_{\mathrm{test},\nu}|\bigr)\cdot \left|\widehat{\mathscr{N}}_p(\lambda,\epsilon)\right|.
    \end{displaymath}
\end{enumerate}
We then set $(\lambda_j^\ast,\epsilon_j^\ast)$ to the pair minimizing the average criterion
$\frac{1}{K}\sum_{\nu=1}^K \mathrm{SCV\text{-}RSS}_\nu(\lambda,\epsilon)$.

To reduce computational cost when the candidate grids are large, we additionally implement a randomized hyperparameter search: rather than evaluating all $(\lambda,\epsilon)$ combinations, a user-defined proportion of pairs is sampled without replacement and assessed via the SCV-RSS criterion.

\setcounter{subsection}{0}
\renewcommand{\thesubsection}{B.\arabic{subsection}}

\subsection{Additional simulation results}\label{sec:add_sim_analysis_res}

We further investigate the finite-sample performance of the proposed approach in a more complex simulation setting. Whereas the main text focused on a single binary covariate to
enable direct comparison with benchmark methods, here we demonstrate the model's
flexibility in handling multiple heterogeneous external variables simultaneously, and use this setting to further discuss sample size requirements. 

\subsubsection{Data generation}
\label{subsubsec:data_gen_supp}

\begin{figure}
\centering
\includegraphics[scale=0.9]{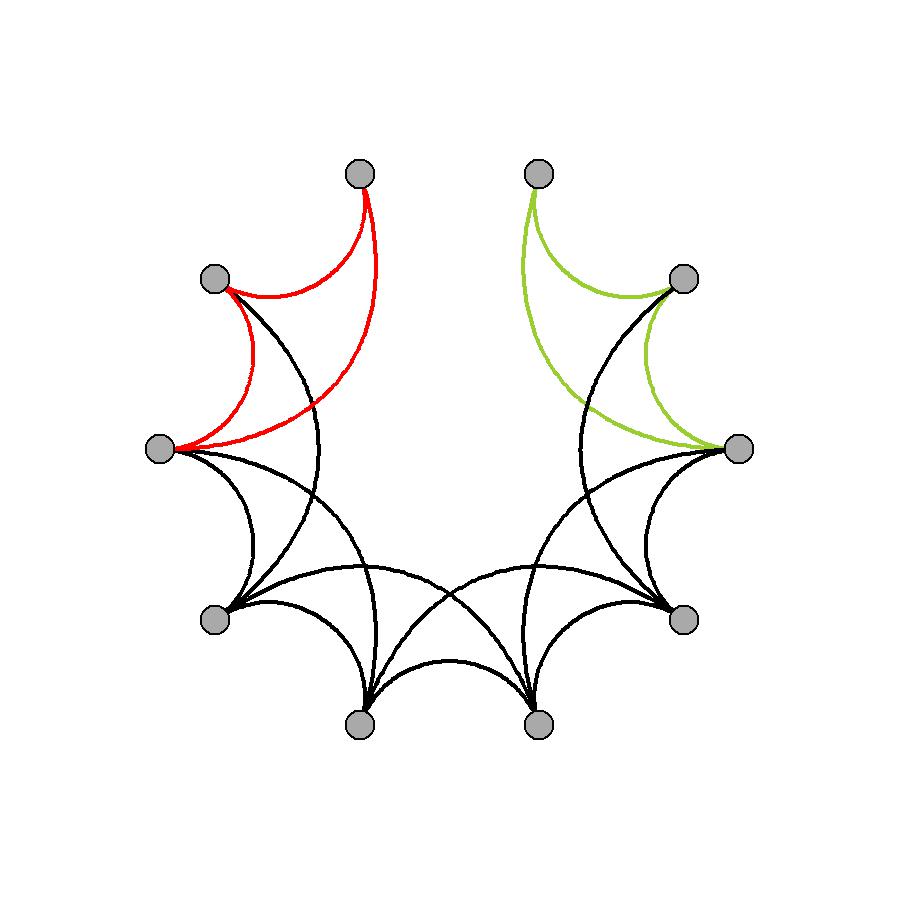}
\caption{Visual illustration of the graphical model for the multiple-covariate simulation
scenario described in Section~\ref{subsubsec:data_gen_supp}. Black solid lines depict
pairwise conditional interdependencies common to both groups; red solid lines depict
pairwise conditional interdependencies present only in $Gr^1$; green lines represent
interdependencies common to both groups but modulated by the continuous
variable.}\label{fig:graph_sims_supp}
\end{figure}

In this additional simulation, we model two external variables: a binary grouping factor and
a positive continuous variable, mimicking a typical biological covariate such as age.
Hence $q=2$: $x_{i,1} \in \{0,1\}$ is the group indicator, with $x_{i,1}=0$ for
observations in the reference group $Gr^0$ and $x_{i,1}=1$ for those in $Gr^1$, with
group sizes balanced so that half the sample belongs to each group; $x_{i,2} \in [0,1]$
is a normalized continuous variable drawn independently of group membership from a
uniform distribution on $[0,1]$.\\
Data are simulated following the procedure described in Section~\ref{sec:data_gen}, where the score vector $(\boldsymbol{\alpha}_{i,1}^{\top},
\dots, \boldsymbol{\alpha}_{i,p}^{\top})^{\top} \in \mathbb{R}^{pM^*}$ is drawn from a
multivariate Gaussian distribution with zero mean and covariance matrix
$\mathbf{\Sigma}_i = (\mathbf{\Theta}^0 + \sum_{q=1}^2 x_{i,q} \mathbf{\Theta}^q)^{-1}$,
where $\mathbf{\Theta}^0$ and $\mathbf{\Theta}^q$ ($q=1,2$) are symmetric matrices of
size $pM^* \times pM^*$ encoding the population network $G^0$, the group-modulated
differential graph $G^1$, and the continuously modulated differential graph $G^2$,
respectively. The precision matrices are constructed so as to simulate a scenario, schematically depicted in Fig.~\ref{fig:graph_sims_supp} where the samples in $Gr^1$ show a set of pairwise conditionally interdependencies not present in the samples within $Gr^0$ while the continuos variable modulates another disjoint set of interdependencies. To this end, similarly to what done in Section~\ref{sec:data_gen} we modify the block-banded matrix in Eq.~(\ref{eq:theta_structure}) by setting $\mathbf{\Theta}^0_{kj} = \mathbf{\Theta}^2_{kj} = 0$ for all
    $k,j \in \{1, \dots, p-\lfloor \frac{p}{3} \rfloor \}$, to simulate the set of edges present only in the graph associated to $Gr^1$, and $\mathbf{\Theta}^0_{kj} = (\mathbf{\Theta}^0 + \mathbf{\Theta}^1)_{kj} = 0$
    for all $k,j \in \{p-\lfloor \frac{p}{3} \rfloor+1, \dots, p \}$, to simulate the set of interdependencies modulated by the continuos variable. We observe that, in this setting, when $x_{i,2}=0$ such interdependencies are not present, while for $x_{i,2}=1$ their strength is comparable to that of the remaining conditional interdependencies in the network.
    
\subsubsection{Results}

\begin{figure}
\centering
\includegraphics[width=\textwidth]{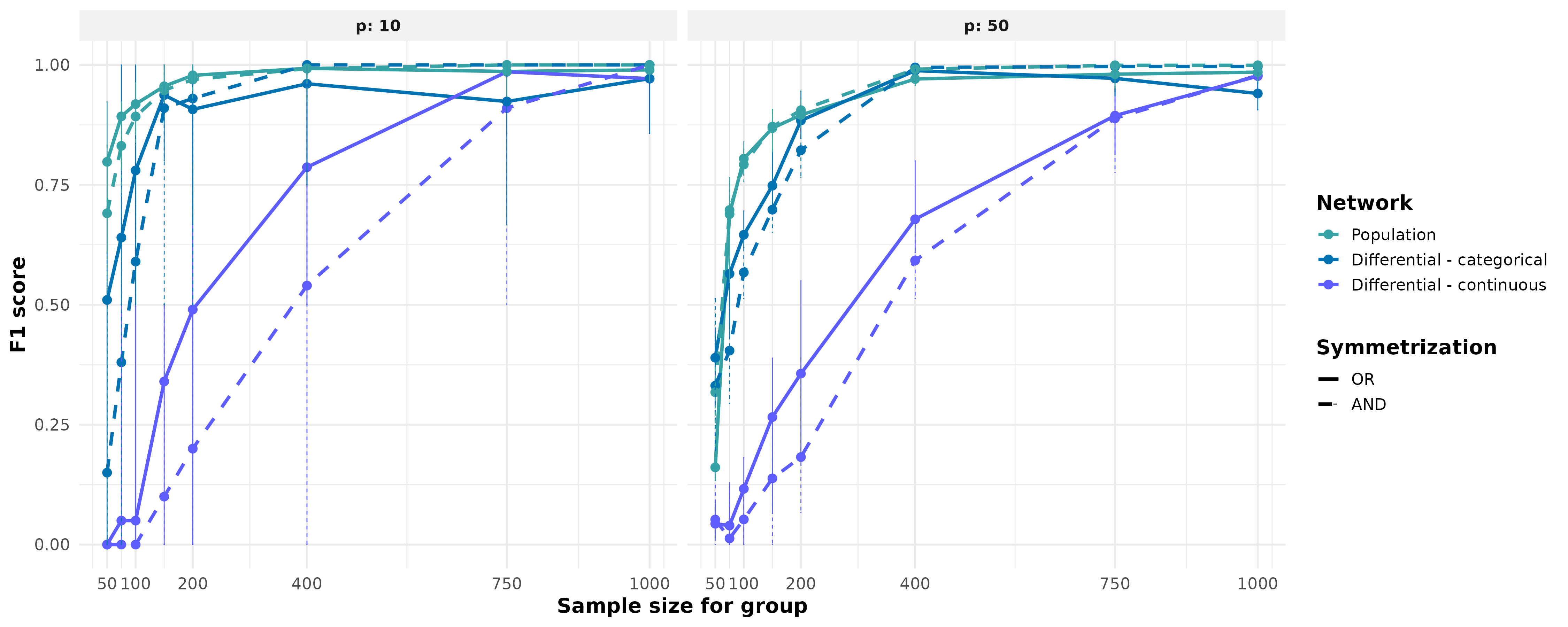}
\caption{Performance in estimating the population network $G^0$, the group-related
differential graph $G^1$, and the differential graph related to the continuos variable $G^2$, for different numbers of nodes $p$, sample sizes $n$, and symmetrization strategies. For each
combination of $p$ and $n$, plots show the minimum, mean, and maximum $F_1$ scores across
10 simulated datasets.}\label{fig:F1score_2_cov_setting}
\end{figure}

For each combination of $p \in \{10,50\}$ and
$n \in 2 \cdot \{50, 75, 100, 150, 200, 400, 750, 1000\}$, we generated 10 simulated
datasets following the design in Section~\ref{subsubsec:data_gen_supp}. The performance in recovering the population network $G^0$ and the differential networks $G^1$ and $G^2$ where assessed by computing the $F_1$ score as described in Section~\ref{subsec:benchmark_and_eval_metrics}. Sample sizes were extended beyond those used in the
main text to better characterize performance in this more complex setting with a larger
number of parameters to estimate. Consistent with the main text results,
Fig.~\ref{fig:F1score_2_cov_setting} confirms that the \texttt{OR} symmetrization strategy in Eq.~(\ref{eq:or_symm}) should be preferred over the \texttt{AND} strategy in Eq.~(\ref{eq:and_symm}) particularly in lower-dimensional settings ($n<200$). 
Furthermore,
Fig.~\ref{fig:F1score_2_cov_setting} shows that estimates of the population network $G^0$
are generally more accurate than those of the differential networks, with the $F_1$ score being
higher for the differential graph $G^1$ associated to the categorical covariate than for the continuously modulated network $G^2$, especially at smaller sample sizes. With sufficient data, the approach provides
accurate estimates of all three networks: when the group sample size $n/2$ exceeds 400,
the average $F_1$ score exceeds 0.70 for $G^2$ and approaches 1 for both $G^0$ and $G^1$.

\begin{figure}
\centering
\includegraphics[width=\textwidth]{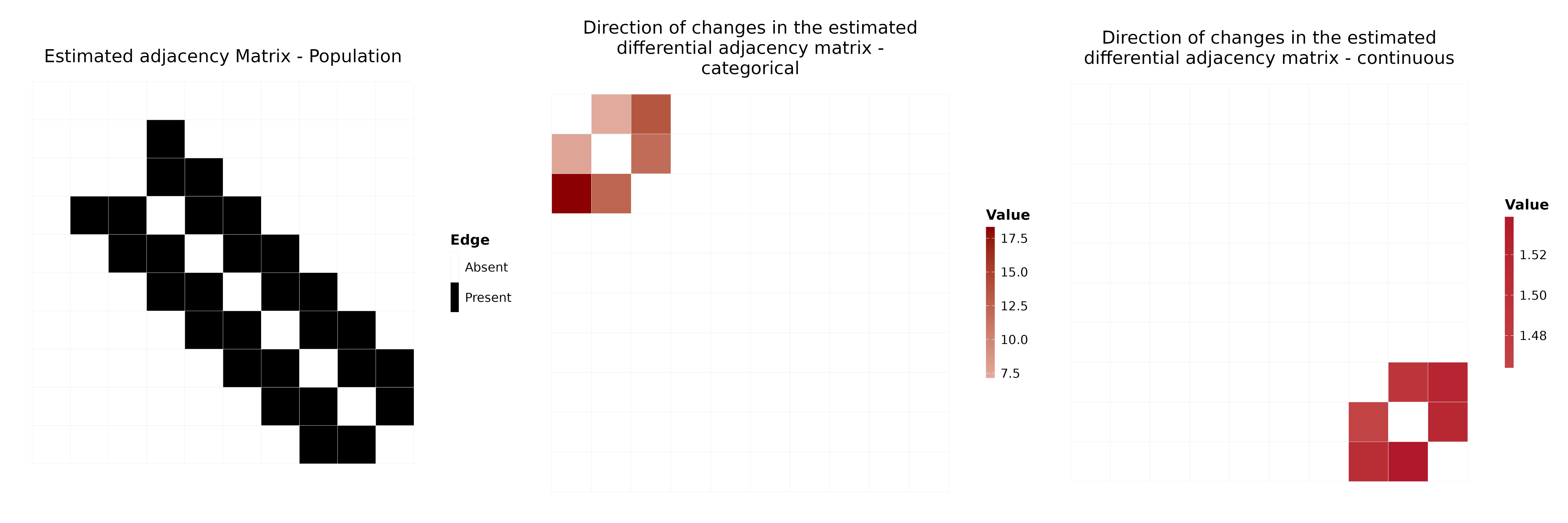}
\caption{Estimated adjacency structure for one simulation replicate with $p=10$ and
$n= 2 \cdot 1000$: population matrix $G^0$ (left), group-related differential matrix
$G^1$ (center), and a continuously related differential matrix $G^2$ (right). Binary edges
are shown for $G^0$ (black = present, white = absent). For the differential matrices,
edges are coloured based on the Variation Rate in Eq.~(\ref{eq:relative_eff}): red indicates a
strengthening and blue an attenuation of the pairwise conditional interdependency relative
to $G^0$.}\label{fig:Supp_res_dir_of_changes}
\end{figure}

Figure~\ref{fig:Supp_res_dir_of_changes} illustrates the pipeline output for a single
replicate with $p=10$ and $n= 2 \cdot 1000$. The pipeline returns the population
graphical model $G^0$, characterizing the mean conditional interdependency structure across
the entire population, the differential graphical model $G^1$, describing how
interdependency strength varies with the grouping factor, and the differential graphical
model $G^2$, describing how interdependency strength varies with the continuous covariate.
Since the algorithm standardizes continuous covariates prior to estimation (Step~1 of
Algorithm~\ref{alg:pipeline}), the population network $G^0$ corresponds to the
connectivity structure at the mean value of $x_{i,2}$ within the reference group $Gr^0$. Edges are coloured according to the Variation Rate in Eq.~(\ref{eq:relative_eff}) showing that the algorithm correctly captures the direction of change. In particular, the differential network $G^1$ captures the additional conditional interdependencies in $Gr^1$ relative to the reference group $Gr^0$,
while the differential network $G^2$ reflects a positive association between the continuous variable and the strength of the interactions among the last subset of nodes.

\subsubsection{Sample size requirements}
\label{subsubsec:sample_size_req}

\begin{figure}
\centering
\includegraphics[width=\textwidth]{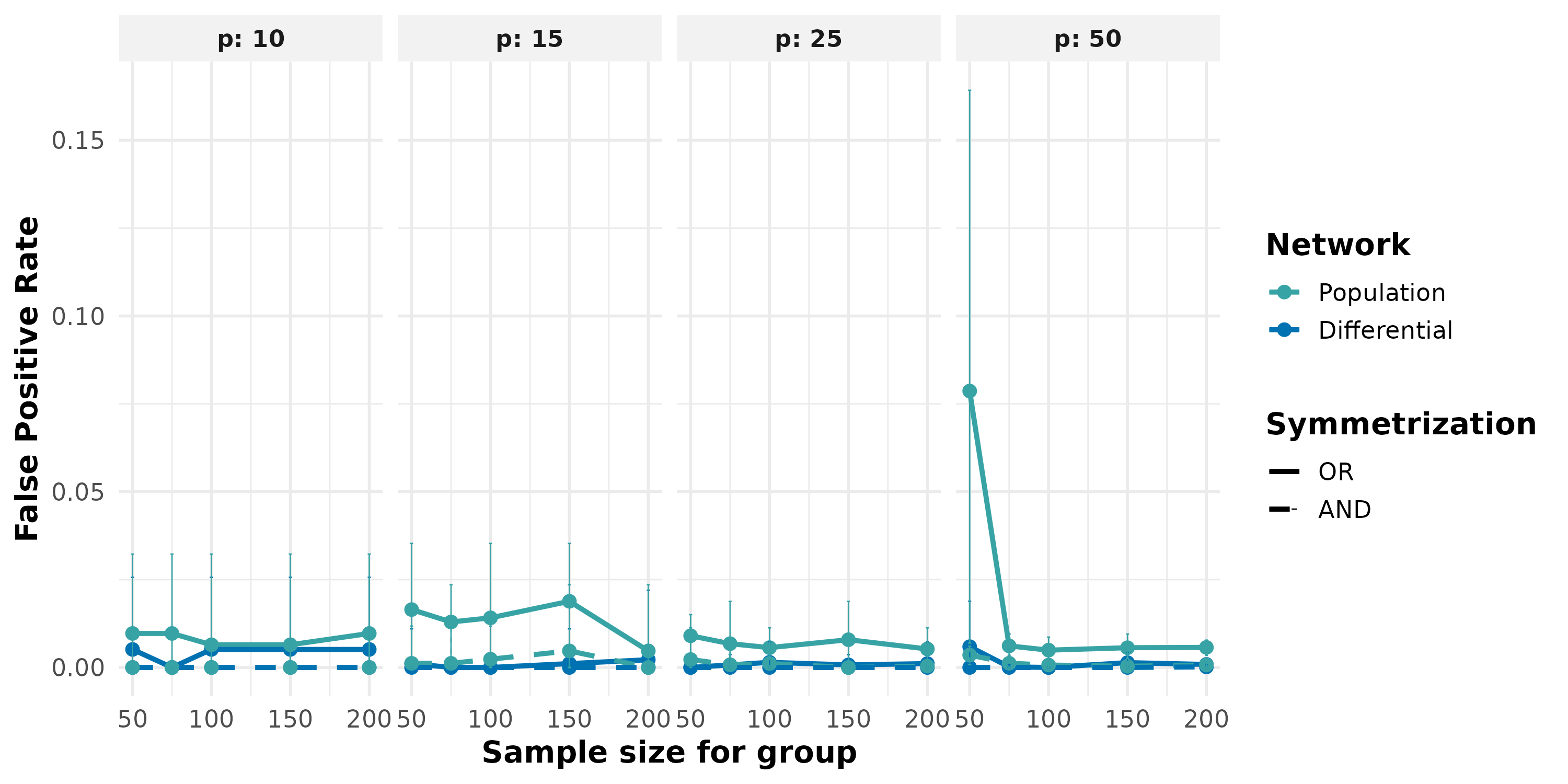}
\caption{False positive rate in estimating the population network $G^0$ and the
differential graph $G^1$ for different numbers of nodes $p$, sample sizes $n$, and
symmetrization strategies, in the single-covariate simulation setting (scenario $S3$ described in Section~\ref{sec:data_gen}). For each combination of $p$ and $n$, plots show the minimum, mean, and maximum
FPR across 10 simulated datasets.}\label{fig:FPRscore_1_cov_setting}
\end{figure}

\begin{figure}
\centering
\includegraphics[width=\textwidth]{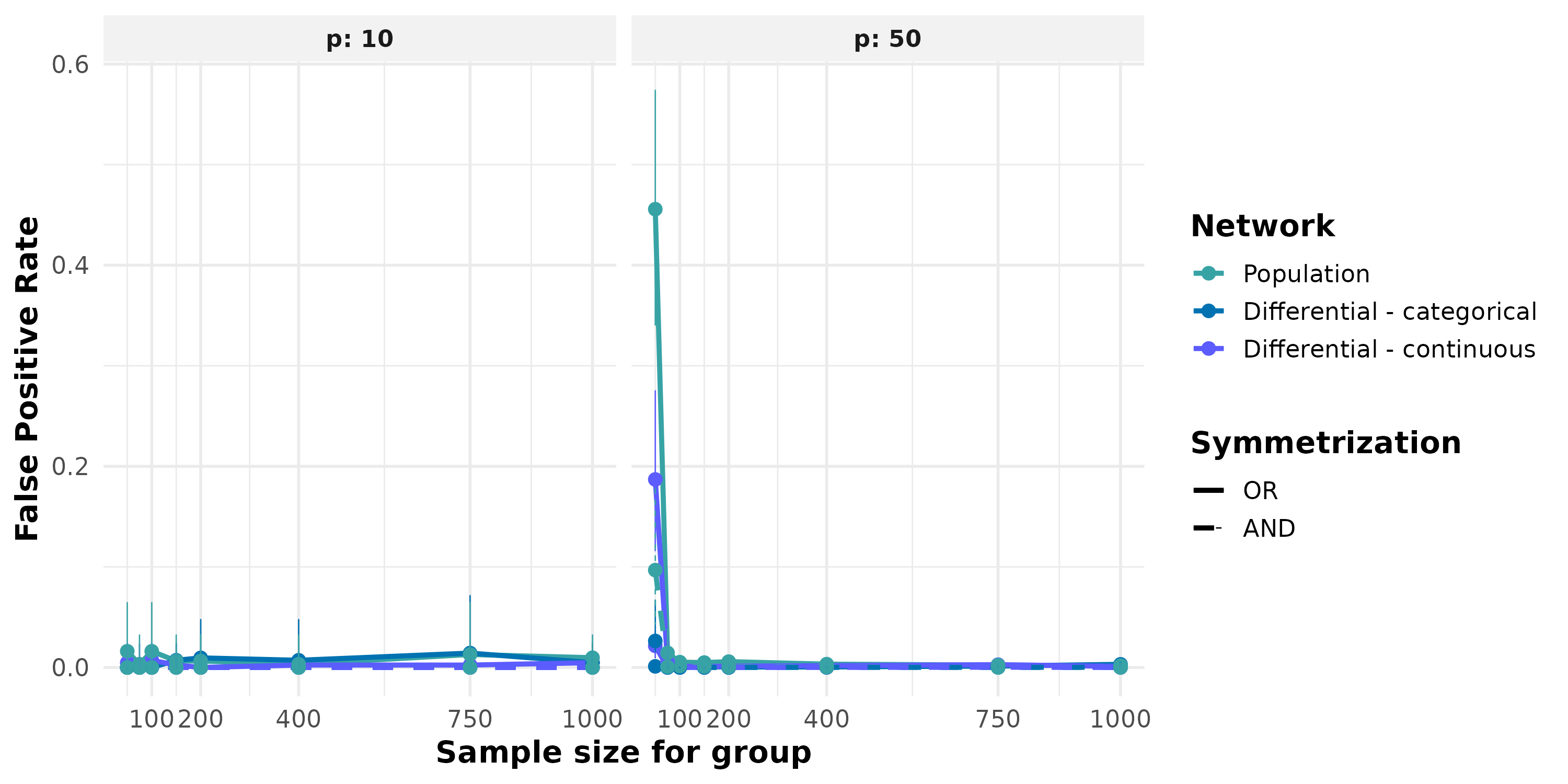}
\caption{False positive rate in estimating the population network $G^0$, the
group-related differential graph $G^1$, and the continuously related differential
graph $G^2$, for different numbers of nodes $p$, sample sizes $n$, and symmetrization
strategies, in the two-covariate simulation setting (described in Section~\ref{subsubsec:data_gen_supp}). For each
combination of $p$ and $n$, plots show the minimum, mean, and maximum FPR across 10
simulated datasets.}\label{fig:FPRscore_2_cov_setting}
\end{figure}

To assess the specificity of the estimated networks, Figs.~\ref{fig:FPRscore_1_cov_setting}
and~\ref{fig:FPRscore_2_cov_setting} report the false positive rate (FPR) in the
single-covariate (scenario $S3$ described in Section~\ref{sec:data_gen}) and two-covariate simulation settings (described in Section~\ref{subsubsec:data_gen_supp}), respectively. In both
settings, the FPR remains at or near zero across nearly all configurations of $p$ and $n$,
with the only exception arising in the limit case of $p=50$ with group sample size
$n/2=50$.


These results indicate that the estimator is conservative: the recovered networks tend to be subgraphs of the true underlying structure, so that even in low sample size settings the identified differential edges can be interpreted as reflecting genuine pairwise conditional interdependencies. As $n$ increases, the model progressively recovers additional true differential edges.

Comparing
Figs.~\ref{fig:F1score_nostro} and~\ref{fig:F1score_2_cov_setting}, the population
network $G^0$ reconstruction is generally stable across configurations, with recovery performance remaining largely consistent whether one or two covariates are included. In contrast, the group-modulated differential network $G^1$ is recovered at lower sample sizes when only one binary covariate is modelled; adding a second covariate increases the total number of parameters to be estimated, requiring larger samples for comparable $F_1$
performance. Furthermore, Fig.~\ref{fig:F1score_2_cov_setting} shows that the differential network $G^2$ requires larger sample sizes than $G^1$. This is expected as the variation in interdependency strength related to the continuous variable is smaller in magnitude than the interdependency difference between $Gr^0$ and $Gr^1$. Similarly, when considering only one binary grouping covariate the lowest $F_1$ score was reached in Scenario S6 (see Fig. \ref{fig:f1_diff_scenarios}) where the two groups shared the same edge support and differed only in interdependency strength. This results in a covariate-related variation of smaller magnitude than in other scenarios, making it harder to identify reliably.

\setcounter{subsection}{0}
\renewcommand{\thesubsection}{C.\arabic{subsection}}

\subsection{Complete output for the case-study on EEG functional connectivity.}\label{sec:whole_output_EEG}

In the case study described in Section \ref{sec:results_experdata}, we apply our method to a set of EEG time-series from two groups of individuals (controls vs AUD patients) considering the group membership as covariate. As detailed in Algorithm \ref{alg:pipeline}, in this scenario our pipeline returns the population graphical model $G^0$, characterizing functional connectivity in the controls, and the differential graphical model $G^1$, describing modulation in connectivity induced by AUD. Furthermore, a weight is associated to each edge in $G^1$
to discriminate impaired or strengthened connections. 

\begin{figure}[h!]
    \centering
\includegraphics[width=0.45\linewidth]{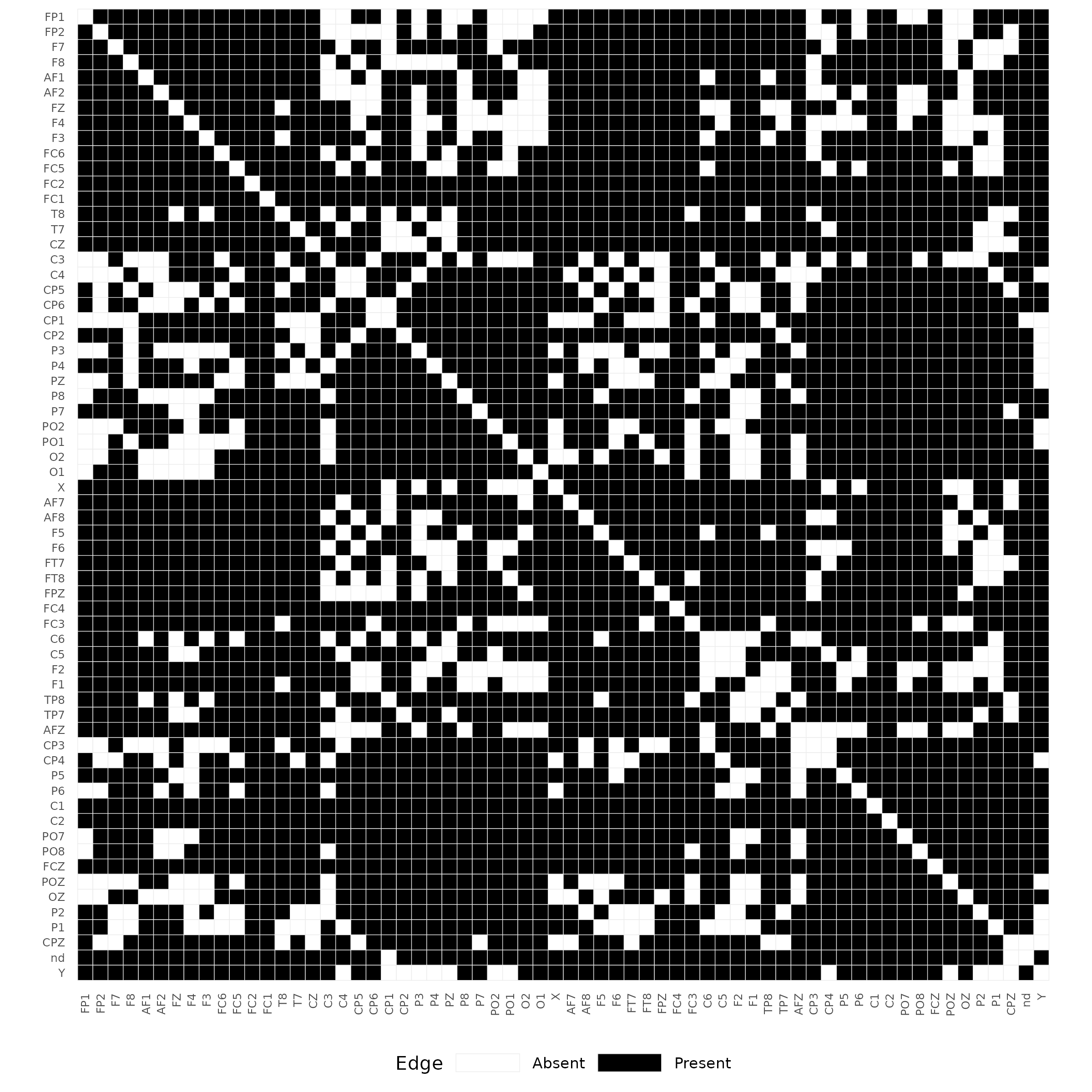}
\includegraphics[width=0.45\linewidth]{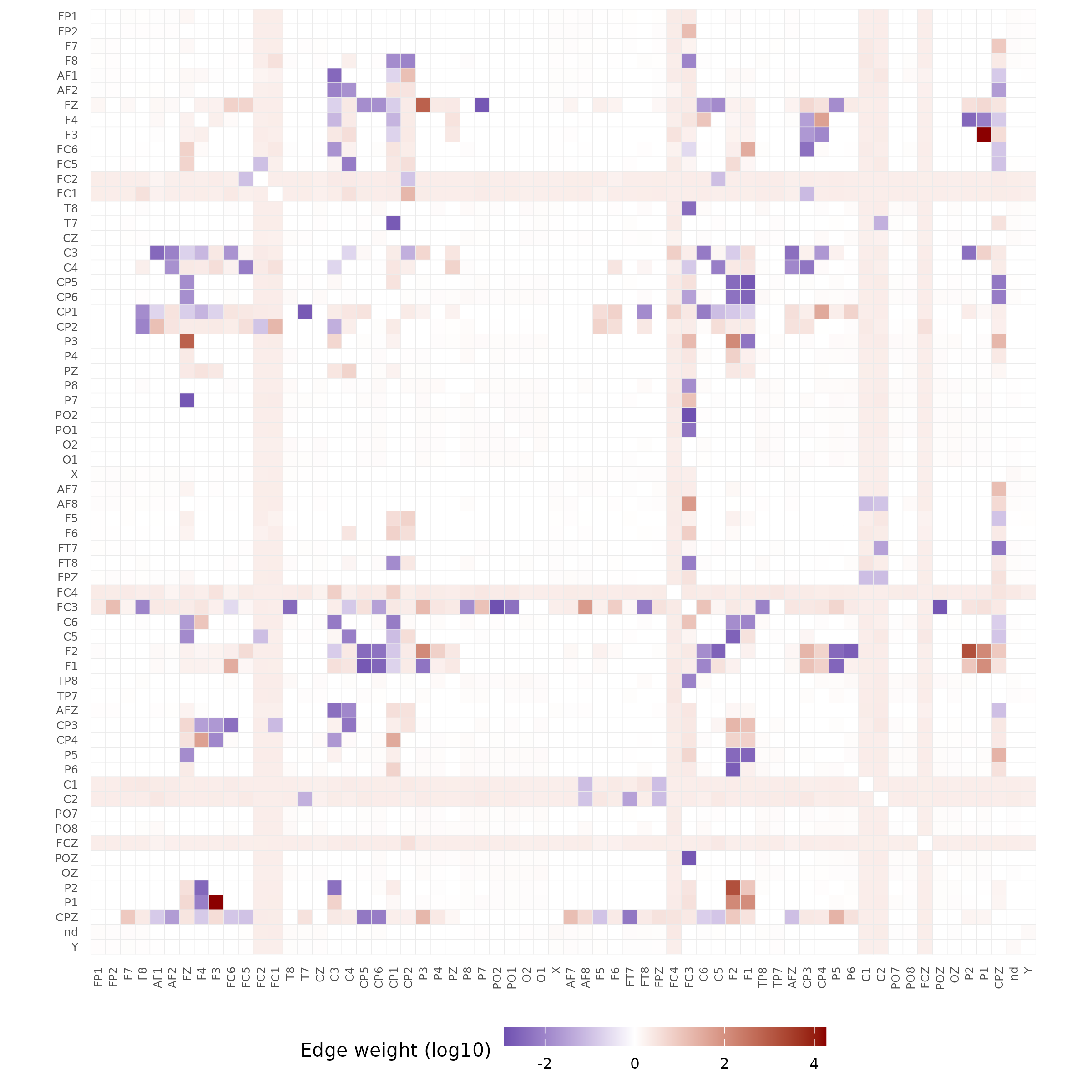}
    \caption{Adjacency matrices describing the population $G^0$ (left) and differential $G^1$(right) functional connectivity in individuals with AUD with respect to controls. For the population matrix, we show binary edges, with black indicating the presence of a connection and white its absence. For the differential matrix, edges are colored according to the weights defined in Eq. (\ref{eq:edge_weights}) in log scale, with blue and red edges indicating reductions and increases in connection strength, respectively.}
    \label{fig:whole_output}
\end{figure}

Fig. \ref{fig:whole_output} displays the adjacency matrices defined by $G^0$ and $G^1$. We observe that the adjacency matrix of $G^0$ shows a block structures indicating inter-hemisphere connectivity mainly involving the frontal areas and the posterior-occipital ones. $G^1$ tends to have a sparser support than $G^0$, which is the typical scenario where directly estimating the differential network may be beneficial \citep{zhao2022}. Further comments on the differential network are provided in Section \ref{sec:results_experdata}.

\bibliographystyle{agsm}
\bibliography{bibliography}

@Article{Boyd2011,
  author  = {Stephen P. Boyd and Neal Parikh and Eric Chu and Borja Peleato and Jonathan Eckstein},
  journal = {Foundations and Trends in Machine Learning},
  title   = {Distributed Optimization and Statistical Learning via the Alternating Direction Method of Multipliers},
  year    = {2011},
  number  = {1},
  pages   = {1--122},
  volume  = {3},
}

@article{cao2022,
  title={Brain functional and effective connectivity based on electroencephalography recordings: A review},
  author={Cao, Jun and Zhao, Yifan and Shan, Xiaocai and Wei, Hua-liang and Guo, Yuzhu and Chen, Liangyu and Erkoyuncu, John Ahmet and Sarrigiannis, Ptolemaios Georgios},
  journal={Human brain mapping},
  volume={43},
  number={2},
  pages={860--879},
  year={2022},
  publisher={Wiley Online Library}
}

@article{carini2025,
  title={Network based statistics shows that rem sleep behavior disorder and visual hallucinations increase functional connectivity in early dementia with {L}ewy bodies},
  author={Carini, Laura and Sommariva, Sara and Fam{\`a}, Francesco and Giorgetti, Laura and Mattioli, Pietro and Orso, Beatrice and Mancini, Raffaele and Pardini, Matteo and Piana, Michele and Arnaldi, Dario},
  journal={medRxiv},
  pages={2025--02},
  year={2025},
  publisher={Cold Spring Harbor Laboratory Press}
}

@article{chen2024,
  title={Estimation of graphical models: An overview of selected topics},
  author={Chen, Li-Pang},
  journal={International Statistical Review},
  volume={92},
  number={2},
  pages={194--245},
  year={2024},
  publisher={Wiley Online Library}
}

@article{deBruin2005,
  title={Associations between alcohol intake and brain volumes in male and female moderate drinkers},
  author={de Bruin, Eveline A and Pol, Hilleke E Hulshoff and Bijl, Suzanne and Schnack, Hugo G and Fluitman, Sjoerd and B{\"o}cker, Koen BE and Kenemans, J Leon and Kahn, Ren{\'e} S and Verbaten, Marinus N},
  journal={Alcoholism: Clinical and Experimental Research},
  volume={29},
  number={4},
  pages={656--663},
  year={2005},
  publisher={Wiley Online Library}
}

@article{hasoon2024,
  title={{EEG} functional connectivity differences predict future conversion to dementia in mild cognitive impairment with {L}ewy body or {A}lzheimer disease},
  author={Hasoon, Jahfer and Hamilton, Calum A and Schumacher, Julia and Colloby, Sean and Donaghy, Paul C and Thomas, Alan J and Taylor, John-Paul},
  journal={International journal of geriatric psychiatry},
  volume={39},
  number={9},
  pages={e6138},
  year={2024},
  publisher={Wiley Online Library}
}

@article{herrera2016,
  title={Functional connectivity and quantitative {EEG} in women with alcohol use disorders: a resting-state study},
  author={Herrera-D{\'\i}az, Adianes and Mendoza-Qui{\~n}ones, Ra{\'u}l and Melie-Garcia, Lester and Mart{\'\i}nez-Montes, Eduardo and Sanabria-Diaz, Gretel and Romero-Quintana, Yuniel and Salazar-Guerra, Iraklys and Carballoso-Acosta, Mario and Caballero-Moreno, Antonio},
  journal={Brain topography},
  volume={29},
  number={3},
  pages={368--381},
  year={2016},
  publisher={Springer}
}

@article{ingber1997,
  title = {Statistical mechanics of neocortical interactions: Canonical momenta indicatorsof electroencephalography},
  author = {Ingber, Lester},
  journal = {Phys. Rev. E},
  volume = {55},
  pages = {4578--4593},
  year = {1997},
  publisher = {American Physical Society},
}

@article{khajehpour2019,
  title={Computer-aided classifying and characterizing of methamphetamine use disorder using resting-state {EEG}},
  author={Khajehpour, Hassan and Mohagheghian, Fahimeh and Ekhtiari, Hamed and Makkiabadi, Bahador and Jafari, Amir Homayoun and Eqlimi, Ehsan and Harirchian, Mohammad Hossein},
  journal={Cognitive Neurodynamics},
  volume={13},
  number={6},
  pages={519--530},
  year={2019},
  publisher={Springer}
}

@book{lauritzen1996,
  title={Graphical models},
  author={Lauritzen, Steffen L},
  volume={17},
  year={1996},
  publisher={Oxford University Press}
}

@article{lee2023,
  title={Conditional functional graphical models},
  author={Lee, Kuang-Yao and Ji, Dingjue and Li, Lexin and Constable, Todd and Zhao, Hongyu},
  journal={Journal of the American Statistical Association},
  volume={118},
  number={541},
  pages={257--271},
  year={2023},
  publisher={Taylor \& Francis}
}

@article{liu2014,
  title={Direct learning of sparse changes in Markov networks by density ratio estimation},
  author={Liu, Song and Quinn, John A and Gutmann, Michael U and Suzuki, Taiji and Sugiyama, Masashi},
  journal={Neural computation},
  volume={26},
  number={6},
  pages={1169--1197},
  year={2014},
  publisher={MIT Press}
}

@article{Meinshausen2006,
  title={High-Dimensional Graphs and Variable Selection with the Lasso},
  author={Meinshausen, Nicolai and Bühlmann, Peter},
  journal={The Annals of Statistics},
  volume={34},
  number={3},
  pages={1436-1462},
  year={2006},
  publisher={Institute of Mathematical Statistics}
}

@article{moysidis2021,
  title={Joint functional Gaussian graphical models},
  author={Moysidis, Ilias and Li, Bing},
  journal={arXiv preprint arXiv:2110.06653},
  year={2021}
}

@article{mueller2013,
  title={Individual variability in functional connectivity architecture of the human brain},
  author={Mueller, Sophia and Wang, Danhong and Fox, Michael D and Yeo, BT Thomas and Sepulcre, Jorge and Sabuncu, Mert R and Shafee, Rebecca and Lu, Jie and Liu, Hesheng},
  journal={Neuron},
  volume={77},
  number={3},
  pages={586--595},
  year={2013},
  publisher={Elsevier}
}

@article{mumtaz2018,
  title={An {EEG}-based functional connectivity measure for automatic detection of alcohol use disorder},
  author={Mumtaz, Wajid and Kamel, Nidal and Ali, Syed Saad Azhar and Malik, Aamir Saeed and others},
  journal={Artificial intelligence in medicine},
  volume={84},
  pages={79--89},
  year={2018},
  publisher={Elsevier}
}

@article{nentwich2020,
  title={Functional connectivity of {EEG} is subject-specific, associated with phenotype, and different from {fMRI}},
  author={Nentwich, Maximilian and Ai, Lei and Madsen, Jens and Telesford, Qawi K and Haufe, Stefan and Milham, Michael P and Parra, Lucas C},
  journal={NeuroImage},
  volume={218},
  pages={117001},
  year={2020},
  publisher={Elsevier}
}

@article{qiao2019,
  title={Functional graphical models},
  author={Qiao, Xinghao and Guo, Shaojun and James, Gareth M},
  journal={Journal of the American Statistical Association},
  volume={114},
  number={525},
  pages={211--222},
  year={2019},
  publisher={Taylor \& Francis}
}

@book{ramsay2005,
  title={Functional data analysis},
  author={Ramsay, JO and Silverman, BW},
  year={2005},
  publisher={Springer},
  address={New York}
}

@article{sakkalis2011,
  title={Review of advanced techniques for the estimation of brain connectivity measured with {EEG/MEG}},
  author={Sakkalis, Vangelis},
  journal={Computers in biology and medicine},
  volume={41},
  number={12},
  pages={1110--1117},
  year={2011},
  publisher={Elsevier}
}

@article{sommariva2025,
  title={Cortical parcellation optimized for magnetoencephalography with a clustering technique},
  author={Sommariva, Sara and Subramaniyam, Narayan Puthanmadam and Parkkonen, Lauri},
  journal={Scientific Reports},
  volume={15},
  number={1},
  pages={6404},
  year={2025},
  publisher={Nature Publishing Group UK London}
}

@article{vallarino2020,
  title={On the two-step estimation of the cross-power spectrum for dynamical linear inverse problems},
  author={Vallarino, Elisabetta and Sommariva, Sara and Piana, Michele and Sorrentino, Alberto},
  journal={Inverse Problems},
  volume={36},
  number={4},
  pages={045010},
  year={2020},
  publisher={IOP Publishing}
}

@article{xu2016,
  title={Semiparametric differential graph models},
  author={Xu, Pan and Gu, Quanquan},
  journal={Advances in neural information processing systems},
  volume={29},
  year={2016}
}

@article{zapata2022,
  title={Partial separability and functional graphical models for multivariate Gaussian processes},
  author={Zapata, Javier and Oh, Sang-Yun and Petersen, Alexander},
  journal={Biometrika},
  volume={109},
  number={3},
  pages={665--681},
  year={2022},
  publisher={Oxford University Press}
}

@article{zhao2014,
  title={Direct estimation of differential networks},
  author={Zhao, Sihai Dave and Cai, T Tony and Li, Hongzhe},
  journal={Biometrika},
  volume={101},
  number={2},
  pages={253--268},
  year={2014},
  publisher={Oxford University Press}
}

@article{zhao2022,
  title={{FuDGE}: A method to estimate a functional differential graph in a high-dimensional setting},
  author={Zhao, Boxin and Wang, Y Samuel and Kolar, Mladen},
  journal={Journal of Machine Learning Research},
  volume={23},
  number={82},
  pages={1--82},
  year={2022}
}

@article{zhao2024,
  title={High-dimensional functional graphical model structure learning via neighborhood selection approach},
  author={Zhao, Boxin and Zhai, Percy S and Wang, Y Samuel and Kolar, Mladen},
  journal={Electronic Journal of Statistics},
  volume={18},
  number={1},
  pages={1042--1129},
  year={2024},
  publisher={The Institute of Mathematical Statistics and the Bernoulli Society}
}

@book{ramsay2005functional,
  title={Functional data analysis},
  author={Ramsay, James O and Silverman, Bernard W},
  year={2005},
  publisher={Springer}
}

@article{Knyazev2007,
title = {Motivation, emotion, and their inhibitory control mirrored in brain oscillations},
journal = {Neuroscience \& Biobehavioral Reviews},
volume = {31},
number = {3},
pages = {377-395},
year = {2007},
author = {Gennady G. Knyazev}
}

@article{zhang1995,
  title={Event related potentials during object recognition tasks},
  author={Zhang, Xiao Lei and Begleiter, Henri and Porjesz, Bernice and Wang, Wenyu and Litke, Ann},
  journal={Brain research bulletin},
  volume={38},
  number={6},
  pages={531--538},
  year={1995},
  publisher={Elsevier}
}

@article{Zhu2014,
  title={Bayesian Graphical Models for Multivariate Functional Data},
  author={Hongxiao Zhu and Nate Strawn and David B. Dunson},
  journal={J. Mach. Learn. Res.},
  year={2014},
  volume={17},
  pages={204:1-204:27},
}

@article{bellec2017neuro,
  title={The neuro bureau ADHD-200 preprocessed repository},
  author={Bellec, Pierre and Chu, Carlton and Chouinard-Decorte, Francois and Benhajali, Yassine and Margulies, Daniel S and Craddock, R Cameron},
  journal={Neuroimage},
  volume={144},
  pages={275--286},
  year={2017},
  publisher={Elsevier}
}

\end{document}